\documentclass[paper=letter, fontsize=12pt]{article}

\RequirePackage[T1]{fontenc}
\RequirePackage{amsthm,amsmath,amsfonts}
\usepackage{longtable}
\usepackage{graphicx}
\usepackage{subcaption}
\usepackage{latexsym}
\usepackage{multirow}

\RequirePackage[numbers]{natbib}
\RequirePackage[colorlinks,citecolor=blue,urlcolor=blue]{hyperref}

\allowdisplaybreaks[1]

\usepackage{times}
\usepackage{keyval}
\usepackage{calc}

\usepackage[letterpaper]{geometry}
\usepackage{setspace}
\usepackage{fancyhdr}
 \usepackage{tikz}
  \usepackage{pgfplots}
 \pgfplotsset{compat=1.15}

\usepackage{adjustbox}
\usepackage{booktabs}
\usepackage{caption} %
\usepackage{array}
\usepackage{float}
\usepackage{enumitem} 
\newcommand{\papertitle}{Arbitrage-Free Bond and Yield Curve Forecasting with Neural Filters under {HJM} Constraints} %
\makeatletter %
\@addtoreset{equation}{section}
\makeatother  %
\numberwithin{equation}{section}
\theoremstyle{plain}
\newtheorem{theorem}{Theorem}[section]%

\newtheorem{proposition}[theorem]{Proposition}

\newtheorem{variational inequality}[theorem]{Variational inequality}

\DeclareMathOperator*{\argmin}{arg\,min}
\DeclareMathAlphabet{\pazocal}{OMS}{zplm}{m}{n}
\DeclareSymbolFontAlphabet{\mathrm}    {operators}
\DeclareSymbolFontAlphabet{\mathnormal}{letters}
\DeclareSymbolFontAlphabet{\mathcal}   {symbols}
\DeclareMathAlphabet{\mathbf}{OT1}{cmr}{bx}{n}
\DeclareMathAlphabet{\mathsf}{OT1}{cmss}{m}{n}
\DeclareMathAlphabet{\mathit}{OT1}{cm}{m}{it}
\DeclareMathAlphabet{\mathpzc}{OT1}{pzc}{m}{it}
\graphicspath{{figures/}}

\begin{document}
	\rhead{\textit{Nov. 2025}}
	\lhead{\textit{Gao \& Hyndman}}
	\chead{\textit{Arbitrage-Free Yield Curve and Bond Forecasting}}

	\title{\papertitle}
	
	\author{
		Xiang Gao\footnote{ 
			Department of Mathematics and Statistics, 
			Concordia University, 
			1455 Boulevard de Maisonneuve Ouest,
			Montr\'eal, Qu\'ebec,
			Canada H3G 1M8.
		}
		\quad and \quad
		Cody Hyndman\footnotemark[1]\ \footnote{Corresponding Author: cody.hyndman@concordia.ca}
	}
	
	\date{November 3, 2025}
	
	\maketitle
	
	\abstract{We develop an arbitrage-free deep learning framework for yield curve and bond price forecasting based on the Heath-Jarrow-Morton (HJM) term-structure model and a dynamic Nelson-Siegel parameterization of forward rates. Our approach embeds a no-arbitrage drift restriction into a neural state-space architecture by combining Kalman, extended Kalman, and particle filters with recurrent neural networks (LSTM/CLSTM), and introduces an explicit arbitrage error regularization (AER) term during training. The model is applied to U.S. Treasury and corporate bond data, and its performance is evaluated for both yield-space and price-space predictions at 1-day and 5-day horizons. Empirically, arbitrage regularization leads to its strongest improvements at short maturities, particularly in 5-day-ahead forecasts, increasing market-consistency as measured by bid-ask hit rates and reducing dollar-denominated prediction errors. 
        }

        \vspace{0.2cm}
\noindent \textbf{Keywords:} arbitrage-free modeling, yield curve forecasting, HJM framework, dynamic Nelson--Siegel, Kalman filter, particle filter, neural networks, LSTM, fixed-income term structure.

	\vspace{5mm}
	\noindent
	\textbf{Mathematics Subject Classification (2000):}
	Primary: 91G30; Secondary: 65C30, 60H30

	\renewcommand{\baselinestretch}{1.5}

\section{Introduction}\label{sec:intro}

No-arbitrage modeling remains central to fixed-income analytics, underpinning the pricing of bonds and term structure derivatives as well as risk management. The Heath--Jarrow--Morton (HJM) framework formalizes this by specifying no-arbitrage conditions for the entire forward-rate curve \cite{heath1992bond}. Recent work extends HJM to overnight-rate markets with scheduled jumps, providing tractable arbitrage-free specifications that account for stochastic discontinuities \cite{FontanaGrbacSchmidt2024_Overnight}. In empirical implementations, the Nelson--Siegel (NS) and dynamic Nelson--Siegel (DNS) families \cite{nelson1987parsimonious,diebold2006forecasting} offer parsimonious, interpretable representations widely used for forecasting. At the same time, imposing strict no-arbitrage can sometimes degrade predictive accuracy in empirical settings \cite{christensen2011affine}, highlighting a long-standing tension between economic consistency and out-of-sample performance.

 A key step toward reconciling this tension is the {\bf arbitrage-regularization} framework of \citet{KratsiosHyndman2020Risks}, which learns the \emph{closest} arbitrage-free model to a given factor model within a generalized HJM setting by augmenting the learning objective with an arbitrage penalty. Specializing to term structure models, they derive a tractable penalty and show how to implement it with neural networks, providing a principled machine-learning route to no-arbitrage. Our approach is directly inspired by \citet{KratsiosHyndman2020Risks}: we operationalize arbitrage-regularization as a training penalty in a filter-based deep sequence-forecasting architecture and evaluate horizon- and maturity-specific gains.

Recent advances in machine learning (ML) complement parametric term structure models. Deep architectures have been developed for multi-curve forecasting, including interval/quantile predictions \cite{RichmanScognamiglio2024}, for yield-curve extrapolation at long maturities \cite{AkiyamaMatsuyama2025}, and for DNS-style structures embedded within neural networks \cite{Lee2023DLNS,JoAhnKimJang2025DLNSX}. Parallel work refines DNS with time-varying decay parameters, conditional heteroskedasticity, and macroeconomic factors \cite{CaldeiraEtAl2025JTSA}, while tree-based regime switching layered on DNS captures macroeconomic state dependence \cite{BieDieboldHeLi2024}. To address interpretability, explainable deep models such as LSTM--LagLasso deliver competitive bond-yield forecasts \cite{NunesEtAl2025LSTMLagLasso}. In environments with policy constraints, smooth shadow-rate DNS variants extend DNS to zero-lower bound (ZLB) or effective lower bound (ELB) regimes while preserving its attractive structure \cite{OpschoorVanDerWel2025}. For multi-curve settings more broadly, recent geometric results characterize consistency and conditions for finite-dimensional realizations in HJM \cite{FontanaLanaroMurgoci2025_GeometryMultiCurve}, offering guidance for factor design across curves.

A parallel line of research integrates arbitrage considerations directly into ML objectives. In option pricing and volatility modelling, arbitrage-aware generators produce risk-neutral or arbitrage-free surfaces by design, including risk-neutral generative networks and arbitrage-free volatility generators \cite{XianYanLeungWu2024,VuleticCont2025}. For interest rates, recent risk-neutral autoencoder approaches model the forward-rate manifold under no-arbitrage \cite{LyashenkoMercurioSokol2024AERM,LyashenkoMercurioSokol2024AE_RiskNeutral}. Relatedly, real-world (physical-measure) HJM formulations analyze market viability and local-martingale deflators across multiple term structures \cite{FontanaPlatenTappe2024_RealWorldHJM}, while roll-over risk has been modeled via stochastic control to endogenously generate spreads even outside the classical no-arbitrage paradigm \cite{FontanaPavaranaRunggaldier2023_RollOverRisk}. Looking forward, data-driven (neural) HJM schemes aim at arbitrage-consistent curve generation and forecasting, including scheduled jumps \cite{CuchieroFontanaGnoatto2024_DataDrivenHJM}.

Hybrid filtering--and--learning designs have likewise matured. Classical Kalman, extended Kalman, and particle filters remain effective for sequential state estimation and can be fused with learned components to capture nonlinearities and time variation. Recent neural-augmented filters such as KalmanNet \cite{Revach2022KalmanNet} and Bayesian KalmanNet \cite{DahanEtAl2025BayesianKalmanNet} demonstrate how deep networks can assist or replace parts of the state-space framework while preserving the inductive bias of filtering. This perspective directly motivates our use of filter-based recurrent neural networks (RNNs) to learn dynamic parameters and latent factors while maintaining a transparent state-space backbone.  These results highlight the importance of enforcing financial structure within deep learning architectures for fixed-income forecasting.

\noindent
\paragraph{Contributions:}
We combine filter-based sequential models with deep learning and enforce no-arbitrage via an explicit training penalty, then evaluate horizon- and maturity-specific gains on U.S.\ Treasuries and corporate bonds. Our contributions are:
\begin{itemize}
  \item \textbf{Arbitrage metric and training signal.} We operationalize no-arbitrage via the \emph{Accumulated Excess Return (AER)} penalty $\Lambda(p)$ on a fixed tenor grid, using it both as a training regularizer and an ex post diagnostic.
  \item \textbf{Price-space and yield-space forecasting.} We forecast \emph{prices} directly (EKF/PF) and \emph{yields} (KF), comparing pathways and trade-offs.
  \item \textbf{Architecture for dynamic parameters.} A convolutional-LSTM compresses per-bond panels and a residual LSTM learns time-varying observation noise, enabling end-to-end learning of $(\kappa,\theta,\sigma)$ within a filter-based RNN.
  \item \textbf{Robust errors and stable PF.} We model observation errors with multivariate generalized Gaussian noise and use EKF-assisted importance sampling with systematic resampling.
  \item \textbf{Headline empirical finding.} Arbitrage-regularization delivers the largest gains at the short end (notably 5-day-ahead), materially improving bid--ask hit rates while maintaining competitive MAE/RMSPE.
  \item \textbf{Practical implementation.} Differentiable KF/EKF/PF components are implemented for end-to-end training, with guidance on accuracy--runtime trade-offs.
  \item \textbf{Scope.} In corporate-bond experiments, credit is absorbed into the latent state (rather than modeling a separate spread factor $\xi_t$), focusing on arbitrage-consistent rate dynamics.
\end{itemize}

\paragraph{Organization:} Section~\ref{sec:af_framework} presents the arbitrage-free pricing framework and its DNS realization. Section~\ref{sec:forecasting} details our Kalman, extended Kalman, and particle filter forecasting schemes. Section~\ref{sec:rnn} introduces the deep architecture for dynamic parameterization and arbitrage-aware training. Section~\ref{sec:empirical} reports empirical results and robustness checks; Section~\ref{sec:conclusion} concludes. An appendix contains technical implementation details.

\section{Arbitrage-free pricing framework}\label{sec:af_framework}

 	The no-arbitrage term structure literature builds upon the theoretical structure introduced by \citet{heath1992bond}. \citet{ang2003no} studied affine no-arbitrage term structure models which preclude arbitrage opportunities between the dynamic evolution of the yield curve factors and the yields at different maturity segments. Interest rate forecasting as in \citet{diebold2006forecasting} shows good out-of-sample performance using the no-arbitrage approach. \citet{christensen2011affine} demonstrates that the no-arbitrage approach downgrades the performance when the model is restricted to preclude arbitrage opportunities.  
 	\subsection{HJM forward rate model}

The HJM framework specifies the joint evolution of the entire forward-rate curve so that discounted bond prices are martingales, providing an economically consistent starting point for forecasting and pricing. The Heath-Jarrow-Morton (HJM) model \cite{heath1992bond} provides a powerful framework in modeling  instantaneous forward rates and fixed income assets in an arbitrage-free setting. The theoretical form of the HJM model allows infinite-dimensional combinations of risk factors while finite-dimensional representations or projections lead to implementable models. Given that affine term-structure is widely applied in dynamic models, we consider finite-dimensional affine structure and the arbitrage-free condition under the risk-neutral measure $\mathbb{Q}$
 	\begin{equation}\label{HJM_forward}
 		df\left(t, \tau\right) = \mu\left(t, \tau\right) dt + \sum_{i=1}^d\eta_i\left(t, \tau\right) dW_i(t),
 	\end{equation}
 	where $\tau$ is the tenor from time $t$ to maturity $T$, $W_i(t)$ for $i=1,2,\dots,d$ are independent standard Brownian motions, $\mu\in\mathbb{R}$ is the drift term and $\eta_i\in \mathbb{R}$ for $i=1,2,\dots,d$ are risk factors.  Intuitively, the model represents the entire forward curve as a small set of latent risk factors with loadings that vary by maturity, and rules out arbitrage by ensuring the expected excess return of any zero-coupon bond is zero.  We assume that (\ref{HJM_forward}) is separable in $t$ and $\tau$ and has a finite-dimensional representation by the following affine structure
 	\begin{equation}\label{AF_forward_model}
 		f(t,\tau) = \beta_\tau X_t,
 	\end{equation}
 	for a deterministic loading parameter $\beta_\tau \in\mathbb{R}^{1\times d}$ and a dynamic process $X_t \in\mathbb{R}^{d\times1}$ containing the risk factors. We assume the loading parameter is chosen such that the corresponding yield curves are in the class of Nelson-Siegel term structure models and  risk arises only from the time varying process $X_t$.  
 	
 	Next, we determine the realization of the forward rate process in finite space and the specification of the volatility term.
Finite-dimensional realizations require the drift and volatility to lie in the tangent space of the forward-rate manifold (see Björk and Svensson). Different volatility specifications recover familiar models (e.g., Ho--Lee with constant volatility, Hull--White with exponentially decaying volatility). Empirically,  Principal Component Analysis (PCA) reveals that three factors explain most variation in U.S.\ yields (see, e.g., \cite{litterman1991common}).
The first factor ('level') accounts for 80--90\% of variation; the second ('slope') moves short and long rates in opposite directions and explains most of the remainder; the third ('curvature') captures hump-shaped movements.
Therefore, we consider a three-factor model for $X_t$ with cross-variable interaction instead of independent variables.
For overnight-rate and multi-curve markets, recent HJM extensions incorporate scheduled jumps and multiple term structures while retaining no-arbitrage (e.g., \cite{FontanaGrbacSchmidt2024_Overnight,FontanaLanaroMurgoci2025_GeometryMultiCurve}); our single-curve setup follows the classic HJM tradition but the forecasting ideas carry over.

Calibration of the forward rate model requires that the initial forward curve is based off on empirically observed forward rates. We implement the calibration under a machine learning framework where the observations may be sequentially batched into many subsets.
In the following section, we introduce the loading parameter $\beta_\tau$ in exponential space and specify the risk variable $X_t$ as a mean-reverting process which we test in a later section. We define $X_t$ as extended Va\v{s}\'{i}\v{c}ek process
 	\begin{equation}\label{factor_model}
 		d X_t = \kappa_t \left(\theta_t - X_t\right) dt + \sigma_t d W_t,
 	\end{equation}
 	where $\kappa_t$, $\theta_t$ and $\sigma_t$ are functions which depend on $X_t$. Equation (\ref{factor_model}) is the factor model and the risk factor $X_t$ is the state variable. The dynamics of the forward rate model $f$ defined in ({\ref{AF_forward_model}}) with state variable $X_t$ defined in (\ref{factor_model}) is also mean-reverting process
 	\begin{align*}
 		df(t,\tau) =& -\frac{d\beta_\tau}{d\tau}X_t dt + \beta_\tau dX_t
 		= \bar\kappa(t,\tau)\left(\bar\theta(t,\tau) - X_t\right)dt + \sigma_t \beta_\tau d W_t,
 	\end{align*}
 	where
        $$
 	\bar\kappa(t, \tau) = \left(\beta_\tau\kappa_t + \frac{d\beta_\tau}{d\tau}\right), \ \mbox{and }\
 	\bar\theta(t, \tau) = \left(\beta_\tau\kappa_t + \frac{d\beta_\tau}{d\tau}\right)^{-1}\beta_\tau\kappa_t\theta_t.
 	$$
 	From the affine forward rate specification (\ref{AF_forward_model}) we obtain the short rate
 	$$
 		r\left(t\right) = \beta_0 X_t,
 	$$
 	and the value of zero-coupon bond $PV(t,\tau)$ which pays one dollar at time $T = t + \tau$ is given by
 	\begin{equation}\label{pricing_formula}
 		PV(t,\tau) = \exp\left(-\int_0^\tau f(t,s)ds\right) = \exp\left(-B_\tau X_t\right),
 	\end{equation}
 	where $B_\tau = \int_0^\tau \beta_u du$.

We denote by $\Lambda(t,\tau)$ the instantaneous excess return of a $\tau$-maturity zero-coupon bond over the money-market account. No-arbitrage requires $\Lambda(t,\tau)=0$ for all maturities.
The relative bond price
$$Z\left(t, \tau\right) = {-\exp{\left(\int_0^t r(s)ds\right)}}{PV(t,\tau)}$$
representing the bond's excess value over the risk-free investment follows the dynamics
 	\begin{equation*}
 		{dZ\left(t, \tau\right)} = \Lambda\left(t, \tau\right){Z\left(t, \tau\right)}  dt - \sigma_t B_\tau  {Z\left(t, \tau\right)} dW_t,
 	\end{equation*}
 	where
 $
 		\Lambda\left(t, \tau\right) = \frac{1}{2}B_\tau^{} \Sigma_t^{} B^\top_\tau - B_\tau \kappa_t\left(\theta_t - X_t\right) + \left(\beta_\tau - \beta_0\right) X_t
 $ 
 	and
 $
 		\Sigma_t = \sigma_t \sigma_t^\top.
 $
 	Building on the arbitrage-regularization approach of \citet{KratsiosHyndman2020Risks}, we later penalize deviations of $\Lambda(t,\tau)$ from zero over a fixed tenor grid. Note that $\Lambda\left(t, \tau\right)$ defines the instantaneous excess return on the bond above the risk free rate and \citet{heath1992bond} proves that there exists a unique market price of risk such that the forward rate model is arbitrage-free. Therefore, the condition $\Lambda\left(t, \tau\right)=0$ determines risk neutral pricing measure and precludes arbitrage opportunities. We summarize these facts in the following theorem.
 	\begin{theorem}\label{thm_AR}
 		Suppose the forward rate model has an affine structure given by $f(t,\tau) = \beta_\tau X_t$
 		and a mean-reverting state variable defined by %
 		\begin{equation*}
 			d X_t = \kappa_t \left(\theta_t - X_t\right) dt + \sigma_t d W_t,
 		\end{equation*}
 		where $\beta_\tau\in\mathbb{R}^{1\times d}$, $X_t\in\mathbb{R}^{d\times1}$, 
 		$\kappa_t\left(X_t\right):\mathbb{R}^{d\times1} \rightarrow \mathbb{R}^{d\times d}$, 
 		$\theta_t\left(X_t\right):\mathbb{R}^{d\times1} \rightarrow\mathbb{R}^{d\times 1}$, and
 		$\sigma_t\left(X_t\right):\mathbb{R}^{d\times1} \rightarrow\mathbb{R}^{d\times d}$.
 		If, for all $t\geq0$ and $\tau\geq0$, the equation
 		\begin{equation*}
 		\Lambda(t,\tau) = \frac{1}{2}B_\tau^{} \Sigma_t^{} B^\top_\tau - B_\tau \kappa_t\left(\theta_t - X_t\right) + \left(\beta_\tau - \beta_0\right) X_t = 0
 		\end{equation*}
        holds, then $f(t,\tau)$ is an arbitrage-free forward rate model under risk-neutral measure $\mathbb{Q}$.
 	\end{theorem}
The following is a standard result in stochastic calculus which we use to obtain our discretized model.
 	\begin{proposition}\label{Coro_Xt}
 		Suppose $X_t$ evolves as mean-reverting process with time-dependent parameter given by equation (\ref{factor_model}). Then (\ref{factor_model}) with initial condition $X_0$ has a unique solution $X_t$ given by 
 		\begin{equation*}
 			X_t = e^{-\int_0^t \kappa_u du}X_0 + \int_0^t e^{-\int_u^t \kappa_v dv}\kappa_u \theta_u  du + \int_0^t e^{-\int_u^t \kappa_v dv} \sigma_u dW_u,
 		\end{equation*}
 		where the mean and variance of $X_t$ are given by
 		\begin{align*}
 			\mathbb{E}\left[X_{T} | \mathcal{F}_t \right] &= e^{-\int_{t}^{T} \kappa_u du}X_t + \int_t^{T} e^{-\int_u^{T} \kappa_v dv}\kappa_u \theta_u  du,\\
 			\text{Var}\left[X_{T} | \mathcal{F}_t \right] &= \int_t^{T} e^{-\int_u^{T} \kappa_v dv} \Sigma_u e^{-\int_u^{T} \kappa_v^\top dv} du.
 		\end{align*}
 	\end{proposition}

 	\subsection{Dynamic Nelson-Siegel term structure}
        We specialize the HJM loading to the Nelson--Siegel family, which represents forward and yield curves via three economically interpretable factors---level (long-run rate), slope (short-long differential), and curvature (medium-term hump).  This three-factor structure aligns with classic PCA evidence in U.S. yields, where a parallel shift (level), steepness (slope), and curvature account for the vast majority of variation \cite{litterman1991common}.

        Choosing different loading parameters $\beta_\tau$, we can generate forward rate curves by (\ref{AF_forward_model}) that give different shapes of the term structure. Similar to the prediction framework introduced by \citet{diebold2006macroeconomy,diebold2008global}, where they introduced the dynamic Nelson-Siegel term structure and modeled the factors using auto-regressive processes, we apply dynamic Nelson-Siegel term structure within the framework of the arbitrage-free forward rate model. We define the loading parameter $\beta_\tau$ as a three-dimensional vector basis for some constant $\lambda\in\mathbb{R}^+$ by
 	\begin{equation*}
 		\beta_\tau=\left(\beta_1(\tau),\beta_2(\tau),\beta_3(\tau)\right)= \left(1,\, e^{-\lambda \tau}, \, \lambda \tau e^{-\lambda \tau}\right),
 	\end{equation*}
        
 In this $d = 3$--dimensional parameterization of the forward rate, the factors $X_t = (X_1(t), X_2(t), X_3(t))^\top$ control the long-run level $X_1(t)$, the short-versus-long-rate slope $X_2(t)$, and the localized curvature $X_3(t)$, giving the DNS model its interpretability for policy and risk applications. The Nelson--Siegel term structure space $\mathrm{NS}(\tau)$ is spanned by the exponential-polynomial basis $\beta_\tau$ with decay parameter $\lambda$,
 	\begin{equation*}
 		\mathcal{NS}(\tau) = \text{Span}\left\{\left.\left(1, ~e^{-\lambda \tau},~\lambda \tau e^{-\lambda \tau}\right)\right|\text{for some } \lambda\in\mathbb{R}^+\right\}.
 	\end{equation*}
 	As shown by \citet{bjork2001existence}, as long as the drift and volatility of the forward rate process lie in $\mathcal{NS}(\tau)$, whose tangent space is itself, then the forward rate process will evolve in $\mathcal{NS}(\tau)$. For some three-dimensional state vector $X_t=\left(X_1(t), ~X_{2}(t), ~X_{3}(t)\right)^\top$, the forward rate model $f(t,\tau)\in\mathcal{NS}(\tau)$
 	\begin{equation*}
 		f(t,\tau) = \beta_\tau X_t = X_1(t) + e^{-\lambda \tau}X_2(t) + \lambda \tau e^{-\lambda \tau} X_3(t),
 	\end{equation*}
 	defines the dynamic Nelson-Siegel yield model with the zero-coupon bond yields given by
 	\begin{equation}
          \label{NS_yield_curve}
 		y(t,\tau) = -\frac{\log \text{PV}(t,\tau)}{\tau} %
 		= \frac{B_\tau}{\tau} X_t %
 		= X_1(t) + X_2(t) \left(\frac{1 - e^{-\lambda \tau}}{\lambda\tau}\right) + X_3(t) \left(\frac{1 - e^{-\lambda \tau}}{\lambda\tau} - e^{-\lambda\tau}\right),
 	\end{equation}
 	where
$$
 		B_\tau = \int_0^\tau \beta_u du =\left(\tau, \frac{1 - e^{-\lambda \tau}}{\lambda}, \frac{1 - e^{-\lambda \tau}}{\lambda} - \tau e^{-\lambda\tau}\right)
                $$
                are the factor loadings.
                In empirical studies of the time series of yields, using the above factor loadings, as in \citet{diebold2006forecasting}, avoids the multicollinearity present in the original Nelson and Siegel \cite{nelson1987parsimonious} specification.
 The term structure space can be expanded to include additional loading terms and different decay parameters so that we can also interpret the forward rate process as the Svensson \cite{svensson1994estimating} term structure model with four state variables:
 	\begin{equation*}
 		\mathcal{SV}(\tau) = \text{Span}\left\{\left.\left(1, ~e^{-\lambda_1 \tau},~\lambda_1 \tau e^{-\lambda_1 \tau},~\lambda_2 \tau e^{-\lambda_2 \tau}\right)\right|\text{for some } \lambda_1,\lambda_2\in\mathbb{R}^+\right\}.
 	\end{equation*}
 	Allowing time-varying or multiple decay parameters can improve fit and forecasts, as in \cite{CaldeiraEtAl2025JTSA}), but we retain a constant decay parameter $\lambda$ for parsimony.
        
 	In order to preclude arbitrage opportunities in the dynamic Nelson-Siegel yield model (\ref{NS_yield_curve}) Theorem~\ref{thm_AR} requires $ \Lambda(t,\tau)= 0$ for all $t\geq 0$ and $\tau \geq 0$. Intuitively, $\Lambda(t,\tau)$ is the model-implied instantaneous excess return of a bond over the risk-free benchmark. Because the set of observed maturities changes from day to day, evaluating $\Lambda(t,\tau)$ only at those maturities yields a time-varying objective. We therefore fix a tenor grid from 3 months to 30 years and define the accumulated excess return (AER) penalty as the time--tenor average $p$-norm of $\Lambda(t,\tau)$ over this grid:
        \begin{equation}\label{arbitrage_penalty}
 		\Lambda^{(p)} = \frac{1}{n}\sum_{i=1}^n\left\|\Lambda(t)(t_i,T_j)\right\|_p= \frac{1}{n}\sum_{i=1}^n\left( {\frac{1}{m}\sum_{j=1}^{m}\left|\Lambda(t)(t_i,T_j)\right|^p}\right)^{\frac{1}{p}}.
 	\end{equation}
We use AER both during training as a regularizer and, ex post, as a diagnostic of arbitrage consistency \cite{KratsiosHyndman2020Risks}.
 When $p=2$ the AER quantifies the average over the set of days $t_{i}$  the root mean square (RMS) distance of the objectives (yields or prices) to the arbitrage-free values over the selected maturities $T_j$ for $i=1,\ldots,n$ and $j=1,\ldots,m$.
	
Moving from risk-free bond prices or yields to corporate bond prices or yields requires additional modelling of credit risk.  Although reduced-form credit spreads $\xi_{t}$ can be added explicitly  (e.g., see \citet{ejsing2012liquidity}), in our corporate-bond prediction applications we shall absorb credit into the latent state for tractability. 
That is, since our interest is to study the arbitrage-free pricing and forecasting problem, we simply assume the $d$ risk factors $X_t$ include credit risk when we apply the model to corporate data instead of modeling it separately.  This keeps the focus on arbitrage-consistent rate dynamics while allowing issuer-level heterogeneity to load through $X_{t}$.
 	
 	Next, we consider the application of the affine term structure in pricing coupon bonds.  Assume the coupon bond periodically pays $c_{i}$ at time $\tau_{i}$ up to $m$ total payments and has value $Y(t, \tau)$ given by the arbitrage-free Nelson-Siegel model (\ref{NS_yield_curve})
 	\begin{equation}\label{AR_pricing}
 		\hat{Y}(t, \tau) = \sum_{i=1}^{m} c_{i} e^{-\tau_{i} y\left(\tau_{i}\right)} = \sum_{i=1}^{m} c_{i} e^{-B_{\tau_{i}} X_t}.
 	\end{equation}
 	From equation (\ref{AR_pricing}), we can extract the yield curve and the state variables from observations. The observations that we used are the daily closing bond prices. We choose the coupon bonds whose tenors are greater than 3 months and less than 30 years. The state variable $X_t$ can be extracted by minimizing the mean square error (MSE) between the observation $Y$ and the model value $\hat{Y}$
 	\begin{equation}\label{objective_function}
 		X_t = \argmin_{X_t\in\mathbb{R}^d} \frac{1}{n}\sum_{i=1}^n \left|Y(t,\tau_i) - \hat{Y}(t,\tau_i)\right|^2,
 	\end{equation}
 	using standard methods such as linear estimators with smoothing penalties. This DNS parameterization provides a finite-dimensional HJM realization, enabling direct enforcement of arbitrage constraints via $\Lambda(t,\tau)$.

 	\subsection{Data and estimation result}
\begin{figure}[!htb]
\begin{minipage}[t]{0.45\textwidth}
 		\centering
 		\caption{Treasury yield curves from 2017 to 2019}
 		\hspace*{-0cm}
 		\includegraphics[width=0.95\linewidth]{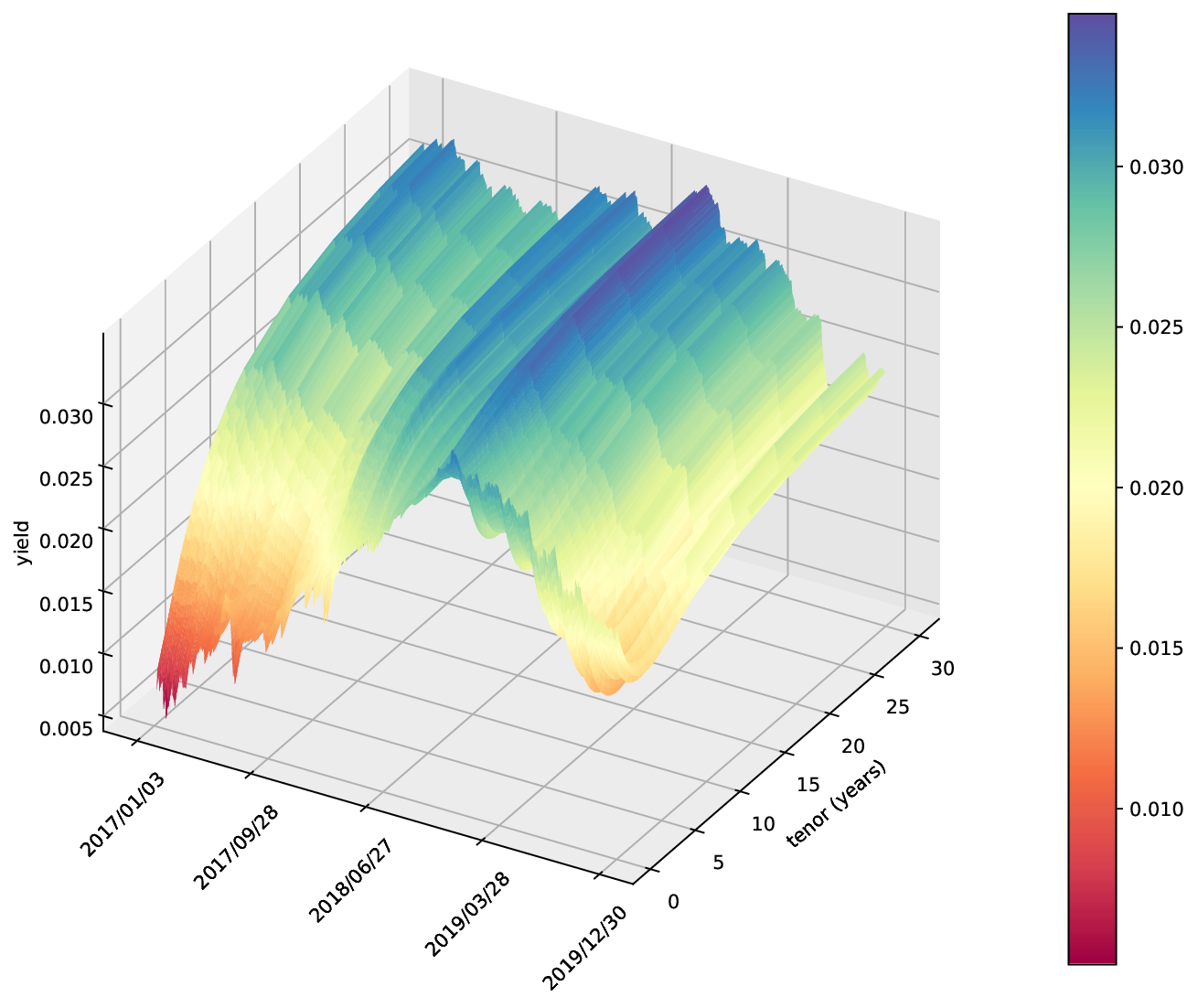}
 		\label{pic1} %
\end{minipage}
\hfill
\begin{minipage}[t]{0.45\textwidth}
 		\centering
 		\caption{State variables of Treasury from 2017 to 2019}
 		\hspace*{-0cm}
 		\includegraphics[width=0.95\linewidth]{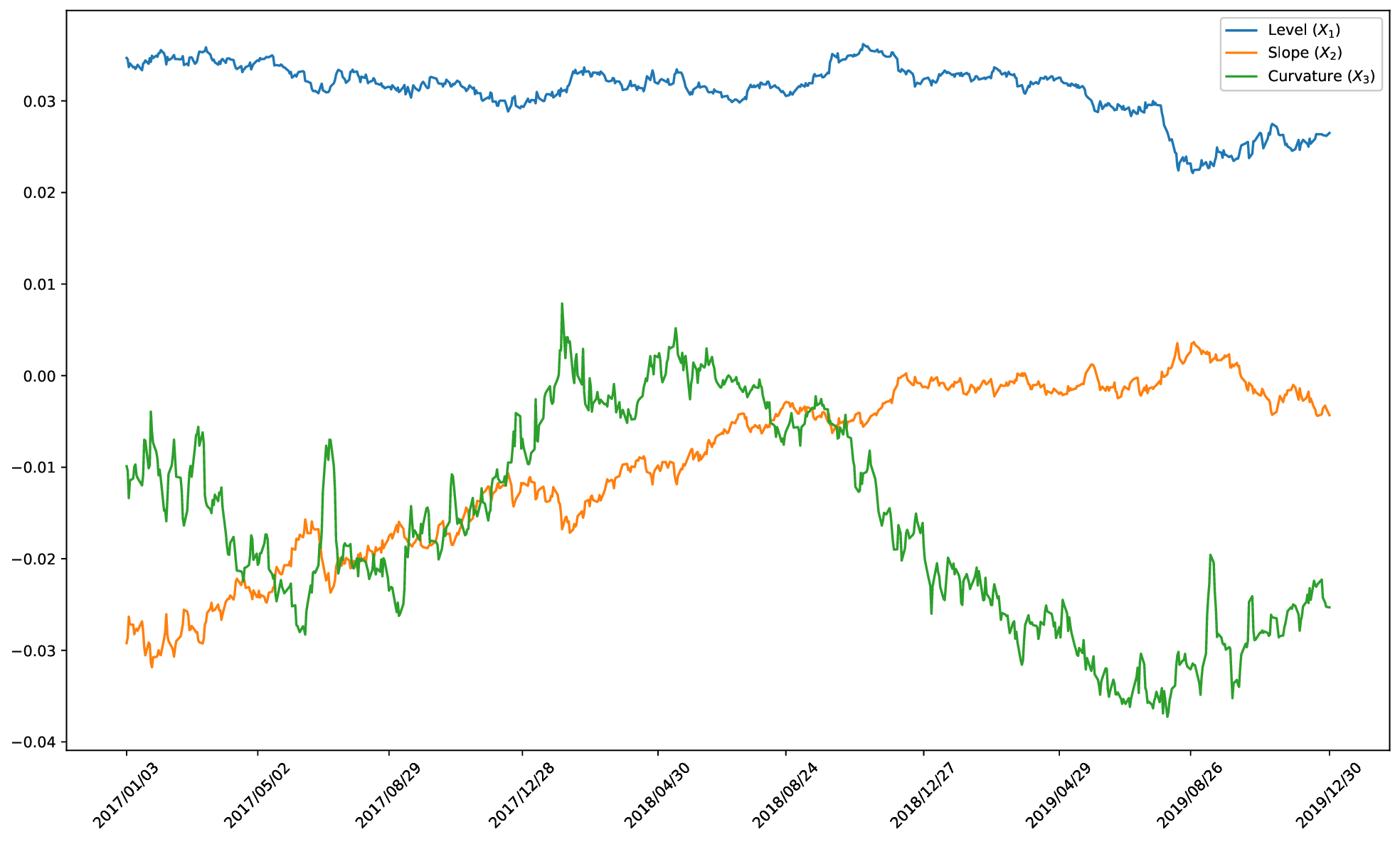}
 		\label{pic2} %
 \end{minipage}
 	\end{figure}

Raw data were obtained from FINRA-TRACE supplemented by proprietary Treasury data feed from a commercial vendor.  Our data comprise daily clean prices, yields to maturity, coupon rates/frequencies, instrument type, convertibility/callability flags, and issue/maturity dates for U.S. Treasuries and corporate bonds from 2017--2019. We retain fixed-coupon, non-callable, non-convertible bonds with remaining time-to-maturity between 3 months and 30 years, yield-to-maturity under 700 bps, and an absolute YTM--coupon difference under 500 bps.      
Because the cross-section varies by day, we sample a fixed panel per day balanced across tenor buckets: approximately 14  short-term (0--2 year) maturity, 45 medium-term (2--10 years), and 9 long-term (10--30 y) Treasuries.  Trading dates with insufficient observations are dropped, amounting to approximately 1--3%
We select the decay parameter $\lambda$ by a grid search minimizing out-of-sample RMSE over a rolling validation window and keep it fixed across the sample for stability.  We then fit the daily coupon bonds using the Nelson-Siegel model (\ref{NS_yield_curve}) to obtain the state variables of $X_i(t)$ for $i=1,2,3$ with the optimal value of decay parameter fixed at $\lambda=0.4488779759$. Figure~\ref{pic1} shows the yield surface and Figure~\ref{pic2} shows the paths of three state variables. 

	\begin{table}[ht]
		\centering
		\caption{U.S. Treasuries yields (in $\%$)}\label{Table00}
			\begin{tabular}{ccccccccccccc}
				\hline
				date & 3M & 6M & 9M & 12M & 15M & 18M & ... & 120M & 180M & 240M & 300M & 360M \\
				\hline
				1/9/2017 & 0.735 & 0.822 & 0.906 & 0.987 & 1.066 & 1.143 & ... & 2.532 & 2.798 & 2.938 & 3.024 & 3.081  \\
				1/10/2017 & 0.648 & 0.745 & 0.839 & 0.929 & 1.016 & 1.100 & ... & 2.541 & 2.806 & 2.946 & 3.031 & 3.088 \\
				1/11/2017 & 0.672 & 0.768 & 0.861 & 0.950 & 1.035 & 1.117 & ... & 2.535 & 2.794 & 2.932 & 3.015 & 3.071 \\
				1/12/2017 & 0.695 & 0.785 & 0.873 & 0.958 & 1.039 & 1.118 & ... & 2.531 & 2.797 & 2.939 & 3.024 & 3.082 \\
				1/13/2017 & 0.702 & 0.791 & 0.879 & 0.963 & 1.045 & 1.124 & ... & 2.547 & 2.817 & 2.960 & 3.047 & 3.105 \\
				...\\
				\hline
			\end{tabular}
	\end{table}

 	\begin{table}[ht]
 		\centering
 		\caption{Statistics of state variables}\label{Table01}
 			\begin{tabular}{cccccccccc}
 				\hline
 				Factor & Mean & Std & Min. & Max. & \multicolumn{3}{c}{Correlation}& ADF & P-value \\
 				\hline
 				$X_1(t)$ & 0.03120 & 0.0030 & 0.0221 & 0.0362 & 1 & -0.541 & 0.430 & -1.601 & 0.483 \\
 				$X_2(t)$ & -0.0098 & 0.0093 & -0.0310 & 0.0041 & - & 1 & -0.353 & -2.345 & 0.157\\
 				$X_3(t)$ & -0.0163 & 0.0111 & -0.0366 & 0.0049 & - & - & 1 & -1.321 & 0.619\\
 				\hline
 			\end{tabular}
 	\end{table}

        Table \ref{Table00} presents the daily yields data extracted from the U.S. Treasuries data set.   Table \ref{Table01} shows the backtested result of the state variables using the Augmented Dickey-Fuller (ADF) test to validate the mean-reversion assumption.  ADF test statistics and p-values do not reject a unit root in $X(t)$, indicating non-stationarity in levels. However, over finite forecasting horizons, locally mean-reverting dynamics remain a reasonable approximation. This assumption is sufficient for the sequential filtering procedures employed in the following sections.

 	\section{Forecasting framework}\label{sec:forecasting}
This section presents three sequential filtering approaches for forecasting bond yields and prices given model parameters $\left(\kappa_t, \theta_t,\sigma_t\right)$. Section~\ref{subsec:Kalman} treats yields with a linear state--space model and the Kalman filter (KF). Sections~\ref{subsec:EKF}-\ref{subsec:PF}  forecast bond prices with nonlinear filters, the extended Kalman filter (EKF) and a particle filter (PF). We then outline how Section~\ref{sec:rnn} endows these parameters with data-driven dynamics via deep networks (see, e.g., \cite{RichmanScognamiglio2024}; \cite{econometrics10020015}).

\subsection{Yield forecasting using the Kalman filter}\label{subsec:Kalman}
We begin with a linear‐Gaussian specification so that yields admit a standard Kalman update, providing a transparent baseline for later nonlinear price-space filters.

 	Consider the yields $y_t=\left(y_1,\cdots,y_m\right)$ at time $t$ observed for fixed tenors $\tau_1,\cdots,\tau_m$. We assume that the noise between the observations and the state model (\ref{NS_yield_curve}) is Gaussian with mean zero and variance $U_t$
 	\begin{equation}\label{yield_eqn}
 		y(t,\tau) = \frac{B_\tau}{\tau} X_t + \epsilon_t,
 	\end{equation}
 	where $\mathbb{E}\left[\epsilon_t\right]=0$ and $\text{Var}\left[\epsilon_t\right]=U_t$. It is difficult to calculate the conditional expectation and variance of $X_t$ directly using Proposition~\ref{Coro_Xt} if the state variable is non-scalar and the parameters $\kappa_t$, $\theta_t$ and $\sigma_t$ are matrices. Therefore, we make a simplification by assuming that the time increment $\Delta t_k = t_{k+1} - t_k$ is constant and $\kappa_t$, $\theta_t$, and $\sigma_t$  are piecewise constant functions for $t\in[t_k,t_{k+1})$ and all $k\geq 0$. By abuse of notation write
          $\kappa_{k}=\kappa_{t_{k}}$, $\theta_{k}=\theta_{t_{k}}$, and $\sigma_{k}=\sigma_{t_{k}}$ for the constant values over the intervals; $\mathcal{F}_{k}=\mathcal{F}_{t_{k}}$ for the observation filtration; and $X_{k}=X_{t_{k}}$ for the discretized state process.  Then, by Proposition~\ref{Coro_Xt}, we obtain the following approximations
 	\begin{align}
 	\mathbb{E}\left[X_{k+1} | \mathcal{F}_k \right] &=e^{-\kappa_k\Delta t}X_k + \left(I - e^{-\kappa_k\Delta t}\right)\theta_k,\label{expectation_eqn}\\
 	\text{Var}\left[X_{k+1} | \mathcal{F}_k \right] &= \int_{t_{k}}^{t_{k+1}} e^{-\kappa_{u}(t_{k+1} - u)}\sigma_{u}\sigma_{u}^\top e^{-\kappa_{u}^{\top}(t_{k+1}-u)}\, du %
               = \int_{t_{k}}^{t_{k+1}} e^{-\kappa_k (t_{k+1}-u)} \Sigma_k e^{- \kappa_k^\top (t_{k+1}-u)} du.
                \label{cov_eqn}
 	\end{align}
 	We denote (\ref{cov_eqn}) as $Q_k=\text{Var}\left[X_{k+1} | \mathcal{F}_k \right]$ and the computation of $Q_k$ can be simplified using the diagonalization of the matrix %
 $		\kappa_k^{} = E^{}_{k} V^{}_{k} E^{-1}_{k}$,
 	where $E_{k}$ is the $(d\times d)$ matrix with the eigenvectors $\kappa_k$, and $V_{k}$ is the diagonal matrix consisting of the $d$ eigenvalues $\zeta^{k}$ of $\kappa_k$. The integral in (\ref{cov_eqn}) can be simplified to
 	\begin{equation}\label{Q_int}
 		Q_k = E^{}_k\left(\int_{t_{k}}^{t_{k+1}} e^{-V^{}_{k}(t_{k+1}-u)} \Omega^{}_{k} e^{-V^\top_{k} (t_{k+1}-u)} dt\right) E^\top_{k} du,
 	\end{equation}
 	where $\Omega^{}_{k} = E^{-1}_{k} \Sigma_k^{} E^{-\top}_{k} = \left(\omega_{i,j}^{k}\right)_{i,j}$. The $(i,j)$-th entry of the integral in (\ref{Q_int}) can be simplified to
 	\begin{align*}
 	I^{k}_{i,j} &=	\int_{t_{k}}^{t_{k+1}} \left(e^{-V^{}_{k}(t_{k+1}-u) } \Omega^{}_{k} e^{-V^\top_{k}(t_{k+1}-u)}\right)_{i,j}\, du
 		= \int_{t_{k}}^{t_{k+1}} e^{-\zeta_{i}^{k}(t_{k+1}-u)}\, (\omega_{i,j}^{k})\, e^{-\zeta_{j}^{k}(t_{k+1}-u)}\, du \\
 		&= \begin{cases}
                  \frac{\omega_{i,j}}{\zeta_{i}^{k} + \zeta_{j}^{k}} \left( 1 - e^{-\left(\zeta_{i}^{k} + \zeta_{j}^{k}\right) \Delta t}\right), & \text{if }
                  \zeta_{i} \neq -\zeta_{j} \\
                  \omega_{i,j}^{k}\Delta t & \mbox{if } \zeta_{i}^{k} = -\zeta_{j}^{k} 
                  \end{cases}.
 	\end{align*}
        In empirical implementations, if $|\zeta_{i}^{k} + \zeta_{j}^{k}|<\epsilon$, for some small tolerance $\epsilon >0$ we would use the second case of the integral approximation.
Alternatively, stable and efficient implementations could employ a Schur decomposition or a Van Loan block exponential to form the conditional variance estimate.
        Nevertheless, in the remainder of this study, we shall estimate $Q_{k}$ by
 	\begin{equation*}
 		Q_k = E^{} \left[\left(\frac{\omega_{i,j}}{\zeta_{i} + \zeta_{j}} \left( 1 - e^{-\left(\zeta_{i} + \zeta_{j}\right) \Delta t}\right) \right)_{i,j}\right] E^{T}.
 	\end{equation*}
	Equations (\ref{yield_eqn}) and (\ref{expectation_eqn}) give the state and observation equations
 	\begin{align*}
 		X_{k+1} = & D_{k} + A_{k}X_{k} + w_k,\\
 		y_{k+1} = & M_{k+1}X_{k+1} + \epsilon_k,
 	\end{align*}
 	where
 	\begin{align*}
 		A_k =& e^{-\kappa_k\Delta t}, \quad
 		D_k = \left(I - e^{-\kappa_k\Delta t}\right)\theta_k, \quad
 		M_{k} = \left[
 		\begin{matrix}
 			\frac{B_{1}\left(\tau_{1}^{k}\right)}{\tau_{1}^{k}},&~\frac{B_{2}^{k}\left(\tau_{1}\right)}{\tau_{1}^{k}},&~\frac{B_{3}\left(\tau_{1}^{k}\right)}{\tau_{1}^{k}}\\
 			\vdots&\vdots&\vdots\\
 			\frac{B_{1}\left(\tau_{m}^{k}\right)}{\tau_{m}^{k}},&~\frac{B_{2}\left(\tau_{m}^{k}\right)}{\tau_{m}^{k}},&~\frac{B_{3}\left(\tau_{m}^{k}\right)}{\tau_{m}^{k}}
 		\end{matrix}\right],
 	\end{align*}
 	and $\tau_m^{k}$ is the maximum tenor from $t_{k}$ among all the observations.
        The noise terms $w_k$ and $\epsilon_k$ are assumed to be independent Gaussian with mean zero and covariance $Q_k$ and $U_k$, respectively.
        The {\it prediction} step of the Kalman filter is given by
 	\begin{align*}
 		\hat{X}_{k|k-1} &= A_{k-1}\hat{X}_{k-1 | k-1} + D_{k-1},\\
 		\hat{P}_{k|k-1} &= A^{}_{k-1} \hat{P}_{k-1|k-1} A^{T}_{k-1} + Q^{}_{k-1},\\
 		\hat{y}_k &= M_{k}\hat{X}_{k \mid k-1},
 	\end{align*}
 	and the {\it update} step is given by
 	\begin{align*}
 		\hat{X}_{k|k} &= \hat{X}_{k|k-1} + K_{k} v_{k},\\
 		\hat{P}_{k|k} &= \hat{P}_{k|k-1} - K_{k} M_{k} \hat{P}_{k|k-1},\\
 		v_{k} &= y_{k} - \hat{y}_k,\\
 		F_{k} &= M_{k} \hat{P}_{k|k-1} M^T_{k} + U_{k-1},\\
 		K_{k} &= \hat{P}_{k|k-1} M^T_{k} F_{k}^{-1}.
 	\end{align*}

 	\subsection{Price forecasting using the extended Kalman filter}\label{subsec:EKF}
To forecast prices directly, rather than yields, we shall replace the linear measurement with a nonlinear bond-pricing map and employ the extended Kalman filter (EKF).
 A related alternative for nonlinear measurement functions is the unscented Kalman filter (UKF), but we do not use it in this paper.

The observations $Y^{}_k=\left(PV_k^{(1)}, \dots, PV_k^{(n)}\right)$ contain $n$ prices of coupon bonds and each observation is defined from (\ref{AR_pricing}) as
 	\begin{equation*}
 		\hat Y(X_t, t) = \sum_{j=1}^{m} c_{\tau_j} e^{-B_{\tau_j} X_t} = C_\tau \exp\left(-B_\tau X_t\right),
 	\end{equation*}
 	where
 	\begin{align*}
 		C_\tau &= \left(c_{\tau_1}, c_{\tau_2},\cdots,c_{\tau_m}\right)\in\mathbb{R}^{1\times m},\\
 		B_\tau &= \left(B_{\tau_1}, B_{\tau_2},\cdots,B_{\tau_m}\right)^T\in\mathbb{R}^{m\times 3}.
 	\end{align*}
 	The extended Kalman filter (see \citet{christensen2011affine}) by the following system
 	\begin{align}\label{EKF1}
 		\hat{X}_{k|k-1} &= A_{k}\hat{X}_{k - 1|k-1} + D_{k},\nonumber\\
 		\hat{P}_{k|k-1} &= A_{k}^{} \hat{P}_{k-1|k-1} A_{k}^T + Q_k,
 	\end{align}
 	and measurement process
 	\begin{align}\label{EKF2}
 		\hat{X}_{k|k} &= \hat{X}_{k|k-1} + K_k v_k,\nonumber\\
 		\hat{P}_{k|k} &= \hat{P}_{k|k-1} - K_k  M_k \hat{P}_{k|k-1},\nonumber\\
 		v_k &= Y_k -  \hat Y(\hat{X}_{k|k-1}, t_k),\\
 		F_k &= M_k \hat{P}_{k|k-1}  M^T_k + U_k,\nonumber\\
 		K_k &= \hat{P}_{k|k-1}  M_k^T F_k^{-1},\nonumber
 	\end{align}
 	where the Jacobian matrix $M_k$ is calculated by
 	\begin{equation*}
 		M_k = \frac{\partial \hat Y(X,t)}{\partial X}\left|_{(\hat{X}_k, t_k)}\right..
 	\end{equation*}
We linearize the measurement function around the current prediction; the Jacobian $M_{k}$ captures local sensitivity.

        Instead of maximizing the log-likelihood, we directly minimize the prediction error
 	\begin{equation*}
 		L(t) = \frac{1}{n} v_k^T v_k,
 	\end{equation*}
and we optionally add the arbitrage penalty (\ref{arbitrage_penalty}) as a regularizer, yielding
        \begin{equation*}
 		L(t) = \frac{1}{n} v_k^T v_k + \lambda \, \Lambda^{(p)}.
 	\end{equation*}

 	\subsection{Price forecasting using the particle filter}\label{subsec:PF}
        Finally, we consider a simulation-based alternative to the EKF, particle filtering, for bond-price prediction. Relative to the EKF, particle filtering (PF) dispenses with functional linearization and Jacobians, accommodating stronger nonlinearities and non-Gaussian errors. The trade-off is higher computation costs due to Monte-Carlo sampling and resampling (PF scales with the particle count; we use systematic resampling when effective sample size falls below a threshold).  Relatedly, \citet{christoffersen2014nonlinear} apply EKF and PF to yield-curve prediction with LIBOR, swap, and cap data.

    In the PF, each state $X(t)=\left(X_1(t),X_2(t),X_3(t)\right)^\top$ of a 3-dimensional vector can be viewed as a particle.      
We use an \textbf{EKF-assisted importance distribution}: for each particle, a local EKF measurement update produces $(\mu_k^{(i)}, P_k^{(i)})$, yielding a Gaussian proposal
$q(X_k \mid X_{k-1}^{(i)}, Y_k) = \mathcal{N}(\mu_k^{(i)}, P_k^{(i)})$,
which lowers weight variance relative to a bootstrap proposal; see \cite{9841276} for a closely related EKF-based importance design. Implementation proceeds by (i) initializing particles from the training-data priors, (ii) propagating via the state dynamics, (iii) updating particle weights using the measurement density, and (iv) systematic resampling when degeneracy is detected. Detailed algorithmic settings (proposal choice, resampling schedule, and effective sample size threshold) are deferred to Appendix~\ref{app:PF}.  We introduce the general sequential Monte Carlo method then we add importance sampling from the measurement equations of the EKF into the PF (see, e.g., \cite{9841276}).

Because price errors exhibit heavy tails and occasional outliers, we model observation noise with a multivariate generalized Gaussian distribution (MGGD), which nests Gaussian and Laplace cases and improves robustness (see, e.g., \cite{pascal2013parameter}).
 	For the prediction errors, we assume a different distribution instead of the multivariate Gaussian. Suppose the marginal densities of observation $Y_t$ given the state $X_t$ can be measured by some distribution $\mathcal{M}$ 
 	\begin{equation*}
 		Y_t\left| X_t \right. \sim \mathcal{M}(Y_t \left| \hat Y\left(X_t, t\right)\right.).
 	\end{equation*}
 	In applications, we shall assume $\mathcal{M}$ is a multivariate generalized Gaussian distribution (MGGD). Following the definition given by \citet{pascal2013parameter} the $n$-dimensional MGGD density is
 	\begin{equation}\label{MGGD}
 		q(x\left|\bar x\right.) = \left|U\right|^{-\frac{1}{2}} C_{p,n} \exp\left(-\frac{\left[\left(x - \bar x\right)^T U^{-1} \left(x - \bar x\right)\right]^p}{2m^p}\right),
 	\end{equation}
 	where $p$ is the shape parameter and $m$ is the scale parameter, $U\in\mathbb{R}^{n\times n}$ is the covariance matrix, and 
 	\begin{equation*}
 		C_{p,n} = \left.p \left(2^{\frac{1}{p}}\pi m\right)^{-\frac{n}{2}} \Gamma\left(\frac{n}{2}\right) \right/\Gamma\left(\frac{n}{2p}\right)
 	\end{equation*}
        is a normalization constant.
 	In particular, if $p = 0.5$, equation (\ref{MGGD}) gives the multivariate Laplace distribution and $p = 1$ gives the multivariate Gaussian distribution. In our model, we treat the MGGD shape and scale parameters as hyperparameters.

Of the three filtering methods the KF offers speed and transparency when a linear yield-measurement is adequate. The EKF enables direct price-space updates but relies on local linearization. The PF handles stronger nonlinearities and heavy-tailed errors at greater computational cost. In Section~\ref{sec:rnn} we show how learned, time-varying $(\kappa_{t},\theta_{t},\sigma_{t})$ further improves all three by aligning parameter dynamics with the data consistent with modern arbitrage-aware term-structure frameworks (see, e.g., \cite{FontanaGrbacSchmidt2024_Overnight}).

	\section{Dynamic parameterization by Recurrent Neural Networks} \label{sec:rnn}
We use a \textbf{filter-aware RNN} composed of: (i) an \emph{input block} that compresses cross-sectional information; (ii) a \emph{state block} that outputs time-varying parameters $(\kappa_t,\theta_t,\sigma_t)$; (iii) a \emph{residual block} that models observation-noise dynamics; and (iv) a differentiable \emph{filter block} (KF/EKF/PF) that closes the loop. The model is trained end-to-end by backpropagating through the filter block.

 	\subsection{Input layer}
 	Since the data are different for the linear model (yield model) and the nonlinear model (price model), we have different input layers.

        \paragraph{Yield-space (linear) model.}
We train on yield panels arranged as a 3D tensor $S\times T\times F$ (samples $\times$ time $\times$ features), where each time step is a $1\times F$ vector. We use yields at $F=23$ fixed tenors $\tau\in\{3,6,\dots,360\}$ months to match the cross-sectional coverage.  We use the extracted yields as inputs and predict the yields as the model output.  To match the proportion of traded bonds in each term bucket there are 8 tenors in the short-term (0-2 year) bucket differing by 3 months
$(3,6,\ldots,24)$, 11 tenors in the mid-term bucket (2-10 year) differing by 6 months between 30 months and 60 months, then by 12 months until 120 months $(30,36,\ldots,60,72,84,\ldots,120)$, and 4 long-term tenors $(180,240,300,360)$ differing by 60 months.

The input layer is a two-layer LSTM that processes the time dimension and outputs hidden states $(c_t,h_t)$; implementation details are given in Appendix~\ref{app:LSTM},  equation (\ref{eq:LSTM}).

\paragraph{Price-space (nonlinear) model.}
The per-day input is an $N\times F$ panel (bonds $\times$ features). We first apply a \emph{convolutional--LSTM (CLSTM)} to compress cross-sectional features into $H$ channels, then feed the result into an LSTM for temporal dynamics. Derivations and kernel-size choices are detailed in Appendix~\ref{app:CLSTM}.  From the input layer, we obtain the final output from the input layer as a vector $c^I_t\in\mathbb{R}^{1\times H}$ and pass it to the state layer.

\subsection{State layer}
 	Suppose we have an output $c^{I}_t\in\mathbb{R}^{1\times H}$ from the input layer and consider it as the input for the state layer. We simply connect the output of the input layer to three dense layers $\kappa$, $\theta$ and $\sigma$ in the state layer
 	\begin{equation*}
 		\begin{alignedat}{2}
 			\kappa(c^{I}_t) :[0,T]\times\mathbb{R}^{1\times H} &\rightarrow \mathbb{R}^{d\times d},~&&\kappa = a_{\kappa}\left(\mathpzc{W}_{\kappa}\cdot c^{I}_t + \mathpzc{b}_{\kappa}\right),\\
 			\theta(c^{I}_t) :[0,T]\times\mathbb{R}^{1\times H} &\rightarrow \mathbb{R}^{d},&&\theta = a_{\theta}\left(\mathpzc{W}_{\theta}\cdot c^{I}_t + \mathpzc{b}_{\theta}\right),\\
 			\sigma(c^{I}_t) :[0,T]\times\mathbb{R}^{1\times H} &\rightarrow \mathbb{R}^{d\times d},~&&\sigma = a_{\sigma}\left(\mathpzc{W}_{\sigma}\cdot c^{I}_t + \mathpzc{b}_{\sigma}\right),
 		\end{alignedat}
 	\end{equation*}
 	where the operator $(\cdot)$ is tensor product. The kernels are
$\mathpzc{W}_{\kappa}\in \mathbb{R}^{H\times d\times d}$, 
$\mathpzc{W}_{\theta}\in \mathbb{R}^{H\times d}$,
$\mathpzc{W}_{\sigma}\in \mathbb{R}^{H\times d\times d}$, 
the biases are
$\mathpzc{b}_{\kappa}\in \mathbb{R}^{d\times d}$,
$\mathpzc{b}_{\theta}\in \mathbb{R}^{d}$,
$\mathpzc{b}_{\sigma}\in \mathbb{R}^{d\times d}$,
and the activation functions are
 $		a_\kappa(x) = x$,
 $ 		a_\theta(x) = \tanh(x)$,
 $ 		a_\sigma(x) = \tanh(x)$.

 \subsection{ Residual layer}
	
 Each time we obtain predicted values $\hat{Y}_t$ we analyze the residual values $e_t = \left|Y_t-\hat{Y}_t\right|$ and estimate the covariance matrix.  We normalize residuals via batch normalization ($BN$), pass them through an LSTM ($L_{R}$), and map the hidden state to $u_{t}$ with a dense layer ($D_{R}$).  The equations in the residual layer are 
 	\begin{align*}
 		\bar e_t =& \text{BN}(e_t), \tag{batch normalization}\\
 		(c^{R}_{t}, h^{R}_t) =& L_R\left(\bar e_t, \left(c^{R}_{t-1}, h^{R}_{t-1}\right)\right), \tag{LSTM}\\
 		u_t =& D_R\left(c^{R}_{t}\right) \tag{dense layer}.
 	\end{align*}

 \subsection{Filter block}       
  
Given $(\kappa_t,\theta_t,\sigma_t,u_t)$, the \emph{filter block} performs a differentiable KF/EKF/PF update to produce state and prediction updates. We train \emph{end-to-end} by backpropagating through this filter block (cf.\ KalmanNet and Bayesian KalmanNet). We obtain the final prediction $\hat{Y}_T$ after feeding the sequential data through the fully connected RNN networks and calculate the values of the arbitrage-free penalties $\Lambda^{(p)}$ using the sequential states $\left(X_{t},\kappa_t,\theta_t,\sigma_t\right)$. The model weights, including all the weights $\mathfrak{W}$ and biases $\mathfrak{b}$ in each layer, will be trained to minimize a weighted sum of prediction error and the arbitrage penalty $\Lambda^{(p)}$.

\begin{figure}[H]
 		\centering
 		\caption{Recurrent Neural Networks}
 		\hspace*{-0.cm}
 		\includegraphics[width=0.75\linewidth, keepaspectratio]{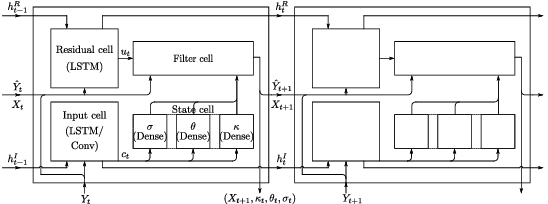}
 		\label{fig6_0_1} %
 	\end{figure}

Each RNN unit comprises these four cells; stacking units across time yields the overall architecture (Figure~\ref{fig6_0_1}). 	

 	\subsection{Objective function}
 Our training objective combines squared prediction error with an arbitrage-regularization term (AER) weighted by $\lambda$:
\[
L(\vartheta)=\frac{1}{n}\sum_{i=1}^n \lVert Y_i-\widehat{Y}_i\rVert^2+\lambda\,\Lambda^{(p)}.
\]
We optimize $L$ via gradient-based learning with early stopping on a rolling validation window.

Section~\ref{sec:empirical} details hyper-parameters, training splits, and ablations (AER on/off, Gaussian vs MGGD, EKF-assisted vs bootstrap PF), and discusses accuracy--runtime trade-offs.

\section{Empirical Results}\label{sec:empirical}

We evaluate our arbitrage-regularized forecasting framework on daily U.S. Treasuries and a panel of corporate bonds (2017--2019), at horizons of \emph{1-day} and \emph{5-day}-ahead. We first describe the dataset and the chronological 80/20 split, then define the evaluation metrics---\emph{MAE} (bps and dollars), \emph{RMSE}, and \emph{MPPE} (percent)---and a bid--ask \emph{hit rate} computed at three spread levels ($\$0.10$, $\$0.25$, $\$0.50$). As a baseline, we reprice bonds using the last {\bf observed} yield curve ({\emph persistence benchmark}). We then compare three filters: (i) \emph{KF} in yield-space (prices via repricing), (ii) \emph{EKF} in price- and yield-space, and (iii) a \emph{PF} in price-space, each trained with and without the AER penalty ($\lambda\in\{0,1\}$). Results are reported separately for Treasuries and corporates, followed by runtime/practicality notes and robustness checks (AER sensitivity, Gaussian vs.\ MGGD errors, EKF-assisted vs.\ bootstrap PF).

\subsection{Data and Splits}
We apply our arbitrage-free prediction models on daily U.S. Treasury bills, notes, and bond data ($\geq 60$  daily observations) and coupon bonds for $12$ corporate bond issuers
($\approx$10-30 daily observations) over the period 2017 to 2019.  Data come from FINRA-TRACE plus a proprietary Treasury feed.  Each observation includes price, tenor, coupon rate, and payment frequency. This uses far fewer features than \citet{ganguli2017machine}, allowing us to isolate the effects of arbitrage-free regularization and other modeling choices.  We use an 80/20 chronological split (2017--2019) with the first 80\% for training/validation and the final 20\% held out for testing.

\subsection{Metrics and Benchmarks}

We organize the data into monthly sequences of $T = 20$ trading days, producing $h$-day-ahead predictions ($h \in \{1,5\}$).  We report MAE (bps) for yields, MAE (dollars) for prices, RMSPE, and MPPE (percentage). The hit rate is computed at three spread levels: \$0.10, \$0.25, and \$0.50 where
\begin{equation}
 		\text{hit rate (spread)} = \frac{1}{NT}\sum_{i=1}^T\sum_{j=1}^N\mathbf{1}_{\left\{\left|Y(t_i,\tau_j) - \hat{Y}(t_i,\tau_j)\right|\leq \text{spread}\right\}}
\end{equation}
is the average over the included tenors and observations for a maturity bucket.  Generally, if the price predictions are within bid-ask spreads we would consider them as market-consistent predictions. To compare the hit rate, we include a {\it persistence benchmark} as a baseline where we hold the last observed yield curve fixed (no-change) and reprice the bonds.   Other measures of prediction accuracy could be used, including volume-weighted or duration-weighted performance measures (e.g., see, \cite{JANKOWITSCH2011343}), but the hit rate defined above provides a crude measure of the degree to which bonds are priced ``correctly'' relative to each other \cite{bliss97}.

For $h=5$, we index trading days $t=0,1,2,\ldots$ (after removing non-trading days) and form five nonoverlapping subsequences ("offsets'')
$\mathcal{T}_r=\{\,t:\ t\equiv r \pmod{5}\,\}$ for $r\in\{0,1,2,3,4\}$.
Each offset contains every fifth trading day (e.g., $t, t+5, t+10,\ldots$), so a time $t$ and its $h$-day-ahead target $t+5$ do not overlap with examples from other offsets; we pool the five offsets for estimation. Offsets are defined by trading-day indices---not calendar weekdays---so they remain valid in holiday weeks.
Separately, to compare forecasting results across maturities, we report yields at 3, 12, 36, 60, 120, 240, and 360 months, and group price results into 0--2 year, 2--10 year, and 10--30 year tenor buckets.  We evaluate KF (yields; prices via bond repricing), EKF (yields and prices), and PF (prices; yields inferred from priced bonds) assisted machine learning models, each with AER on/off ($\lambda\in \{0,1\}$).  
 	\subsection{Main Results (UST)}
 	
 	\begin{table}[ht]
 		\centering
 		\caption{Testing result of U.S. Treasuries: yield prediction error (in bps)}\label{Table1}
 			\begin{tabular}{ccccccc|cccccc}
 				\hline
 				\multicolumn{1}{c} {Model} &\multicolumn{2}{c}{MAPE}&\multicolumn{2}{c}{RMSPE}&\multicolumn{2}{c|}{STDV}&\multicolumn{2}{c}{MAPE}&\multicolumn{2}{c}{RMSPE}&\multicolumn{2}{c}{STDV}\\
 				\cline{2-13}
 				\multicolumn{1}{c} {Maturities}& \multicolumn{1}{c} {1-day} & \multicolumn{1}{c} {5-day } & \multicolumn{1}{c} {1-day} & \multicolumn{1}{c} {5-day } & \multicolumn{1}{c} {1-day} & \multicolumn{1}{c|} {5-day } & \multicolumn{1}{c} {1-day} & \multicolumn{1}{c} {5-day } & \multicolumn{1}{c} {1-day} & \multicolumn{1}{c} {5-day } & \multicolumn{1}{c} {1-day} & \multicolumn{1}{c} {5-day }\\
 				\hline
 				KF ($\lambda$=0) &  &  &  &  &  & & KF ($\lambda$=1) &  &  &  &  &  \\
 				3M & 3.22 & 9.04 & 4.30 & 10.96 & 4.26 & 9.18 & 3.18 & 5.61 & 4.20 & 7.11 & 4.18 & 7.1 \\
 				1Y & 3.17 & 10.3 & 4.20 & 13.25 & 4.17 & 12.07 & 3.23 & 6.3 & 4.26 & 8.19 & 4.22 & 8.13 \\
 				3Y & 3.75 & 12.38 & 4.89 & 16.45 & 4.88 & 16.02 & 3.82 & 8.29 & 5.01 & 11.17 & 4.96 & 11.06 \\
 				5Y & 3.88 & 11.39 & 5.02 & 15.07 & 5.02 & 14.93 & 3.93 & 8.51 & 5.16 & 11.43 & 5.12 & 11.42 \\
 				10Y & 3.72 & 8.49 & 4.73 & 11.13 & 4.73 & 11.11 & 3.85 & 8.29 & 4.92 & 11.14 & 4.91 & 10.83 \\
 				20Y & 3.70 & 9.4 & 4.72 & 12.55 & 4.71 & 12.25 & 3.84 & 9.55 & 4.91 & 12.78 & 4.91 & 11.54 \\
 				30Y & 3.74 & 10.82 & 4.83 & 14.36 & 4.83 & 13.94 & 3.91 & 10.55 & 5.02 & 13.87 & 5.01 & 12.22 \\
 				EKF ($\lambda$=0) &  &  &  &  &  & &EKF ($\lambda$=1) &  &  &  &  &  \\
 				3M & 3.69 & 6.66 & 4.92 & 8.19 & 4.72 & 8.14 & 4.47 & 6.68 & 5.87 & 8.38 & 5.79 & 8.35 \\
 				1Y & 3.40 & 6.53 & 4.57 & 8.27 & 4.48 & 8.27 & 3.88 & 6.64 & 5.19 & 8.41 & 5.14 & 8.33 \\
 				3Y & 3.96 & 8.45 & 5.24 & 11.09 & 5.24 & 11.07 & 4.09 & 8.71 & 5.44 & 11.2 & 5.42 & 11.1 \\
 				5Y & 4.18 & 9.25 & 5.43 & 11.93 & 5.43 & 11.88 & 4.25 & 9.45 & 5.51 & 12.1 & 5.50 & 12.03 \\
 				10Y & 4.03 & 8.91 & 5.11 & 11.44 & 5.11 & 11.36 & 3.98 & 9.18 & 5.07 & 11.78 & 5.06 & 11.77 \\
 				20Y & 3.91 & 8.36 & 4.99 & 11.02 & 4.98 & 10.91 & 3.81 & 8.77 & 4.84 & 11.51 & 4.83 & 11.51 \\
 				30Y & 3.94 & 8.29 & 5.07 & 11.06 & 5.05 & 10.95 & 3.81 & 8.71 & 4.88 & 11.61 & 4.86 & 11.6 \\
 				PF ($\lambda$=0) &  &  &  &  &  & & PF ($\lambda$=1) &  &  &  &  &  \\
 				3M & 4.83 & 8.33 & 6.24 & 10.19 & 6.21 & 9.81 & 4.97 & 7.18 & 6.40 & 9.13 & 6.37 & 9.13 \\
 				1Y & 4.01 & 7.66 & 5.07 & 9.89 & 5.05 & 9.78 & 4.04 & 7.47 & 5.21 & 9.44 & 5.19 & 9.35 \\
 				3Y & 3.97 & 9.23 & 5.15 & 12.06 & 5.15 & 12.06 & 3.94 & 9.6 & 5.10 & 12.08 & 5.09 & 11.86 \\
 				5Y & 4.15 & 9.57 & 5.34 & 12.55 & 5.34 & 12.54 & 4.10 & 10.02 & 5.23 & 12.59 & 5.23 & 12.4 \\
 				10Y & 4.01 & 9.16 & 5.05 & 11.86 & 5.05 & 11.86 & 3.94 & 9.14 & 4.94 & 11.59 & 4.94 & 11.54 \\
 				20Y & 3.93 & 9.07 & 5.01 & 11.88 & 5.00 & 11.84 & 3.85 & 8.33 & 4.85 & 11.02 & 4.84 & 11.02 \\
 				30Y & 4.01 & 9.3 & 5.13 & 12.22 & 5.12 & 12.15 & 3.89 & 8.16 & 4.94 & 11.1 & 4.92 & 11.08 \\
 				\hline
 			\end{tabular}
 	\end{table}
 	
 	\begin{table}[ht]
 		\centering
 		\caption{Testing result of U.S. Treasuries: mean absolute prediction error and mean percentage prediction error}\label{Table1_1}
 			\begin{tabular}{ccccccccccccc}
 				\hline
 				&\multicolumn{2}{c}{\multirow{2}{*}{MAPE (bps)}}&\multicolumn{2}{c}{\multirow{2}{*}{MAPE (dollar)}}&\multicolumn{2}{c}{\multirow{2}{*}{hit rate ($\leq \$0.10$)}}&\multicolumn{6}{c}{MPPE ($\%$) by tenor bucket}\\
 				&&&&&& &\multicolumn{2}{c}{$0\sim2$ (years)}&\multicolumn{2}{c}{$2\sim10$ (years)}&\multicolumn{2}{c}{$10\sim15$ (years)}\\
				& \multicolumn{1}{c}{1-day} & \multicolumn{1}{c} {5-day} & \multicolumn{1}{c}{1-day} & \multicolumn{1}{c} {5-day} & \multicolumn{1}{c}{1-day} & \multicolumn{1}{c} {5-day}& \multicolumn{1}{c}{1-day} & \multicolumn{1}{c} {5-day}& \multicolumn{1}{c}{1-day} & \multicolumn{1}{c} {5-day}& \multicolumn{1}{c}{1-day} & \multicolumn{1}{c} {5-day}\\
 				\hline
 				Benchmark & 3.54 & 6.98 & 0.165 & 0.283 & 54.8$\%$ & 42.58$\%$ & 0.065 & 0.085 & 0.157 & 0.32 & 0.683 & 1.439 \\
 				KF($\lambda$=0) & 3.58 & 10.45 & 0.172 & 0.443 & 54.5$\%$ & 32.66$\%$ & 0.065 & 0.114 & 0.159 & 0.452 & 0.684 & 1.747 \\
 				EKF($\lambda$=0) & 1.95 & 8.11 & 0.181 & 0.354 & 54.1$\%$ & 38.82$\%$ & 0.066 & 0.087 & 0.168 & 0.343 & 0.725 & 1.533 \\
 				PF($\lambda$=0) & 1.95 & 8.85 & 0.180 & 0.373 & 53.5$\%$ & 37.55$\%$ & 0.065 & 0.091 & 0.167 & 0.362 & 0.722 & 1.614 \\
 				KF($\lambda$=1) & 3.67 & 7.96 & 0.175 & 0.363 & 54.4$\%$ & 40.30$\%$ & 0.065 & 0.086 & 0.161 & 0.325 & 0.708 & 1.741 \\
 				EKF($\lambda$=1) & 1.95 & 8.32 & 0.181 & 0.364 & 52.6$\%$ & 38.76$\%$ & 0.065 & 0.09 & 0.168 & 0.399 & 0.741 & 1.673 \\
 				PF($\lambda$=1) & 1.95 & 8.76 & 0.178 & 0.371 & 54.0$\%$ & 36.31$\%$ & 0.067 & 0.096 & 0.157 & 0.365 & 0.693 & 1.58 \\
 				\hline
 			\end{tabular}
 	\end{table}
 	For U.S. treasuries we report the mean absolute prediction error (MAPE), root mean square prediction error (RMSPE) and mean percentage prediction error (MPPE) of the yields (bps) and prices (dollars) in 1-day-ahead and 5-day-ahead predictions for the testing-set in Tables \ref{Table1} and \ref{Table1_1}. Yield forecasting using the dynamic Nelson-Siegel model shows small variation in prediction errors from the short-term maturities to long-term maturities. Yield prediction errors are less than $4.9$ bps in $1$-day-ahead forecasting and less than $11$ bps in $5$-day-ahead forecasting. The price prediction errors are less than $20$ cents in 1-day-ahead forecasting and less than $40$ cents in 5-day ahead forecasting. Arbitrage-free regularization has significant impact on the yield data for the KF-based model but does not influence the price model with EKF-based or PF-based models. It is important to distinguish between yield-space and price-space forecasting: the AER penalty primarily improves yield-space predictions (KF), while having limited influence when forecasting directly in price-space (EKF and PF).  Forecasting performance improvement with arbitrage-free regularization results are presented in Table \ref{Table1}.  For the  KF-based model and 5-day-ahead forecasting horizon the prediction error for 3-month to 5- year maturities are significantly decreased. Forecasting performance of the long-tenor bonds are less accurate than that of the short-tenor bonds.   Forecast errors are larger at long maturities, plausibly reflecting data sparsity; future work will augment the panel with additional long-tenor observations and post-pandemic data.

\begin{figure}[ht]
  \centering
  \caption{U.S. Treasuries: Average Excess Return (\%) by tenor for KF, EKF, PF  1-day-ahead predictions}
  \begin{subfigure}[b]{0.45\textwidth} %
    \centering
  \caption{Training Set AER}
    \includegraphics[width=\textwidth]{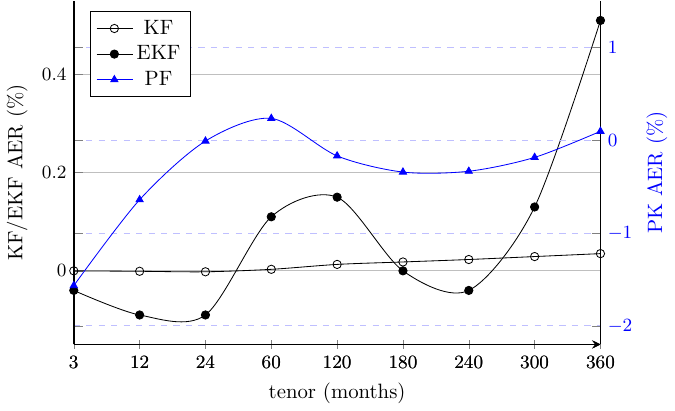} %
    \label{subfig:panel1}
  \end{subfigure}
\hfill
  \begin{subfigure}[b]{0.45\textwidth} %
    \centering
    \caption{Test Set AER}
    \includegraphics[width=\textwidth]{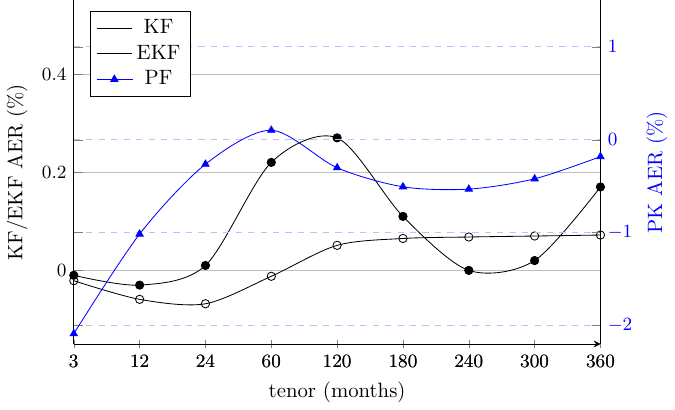} %
    \label{subfig:panel2}
   \end{subfigure}
  \label{fig6_5_1}
\hfill
  \begin{subfigure}[b]{0.45\textwidth} %
    \centering
    \caption{Training Set |AER|}
    \includegraphics[width=\textwidth]{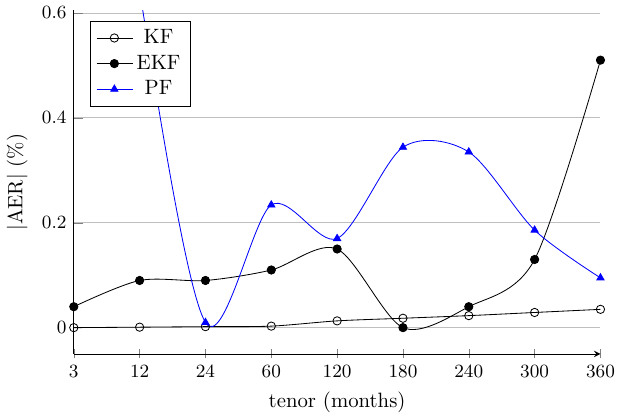} %
    \label{subfig:panel3}
   \end{subfigure}
  \label{fig-aer-3}
\hfill
  \begin{subfigure}[b]{0.45\textwidth} %
    \centering
    \caption{Test Set |AER|}
    \includegraphics[width=\textwidth]{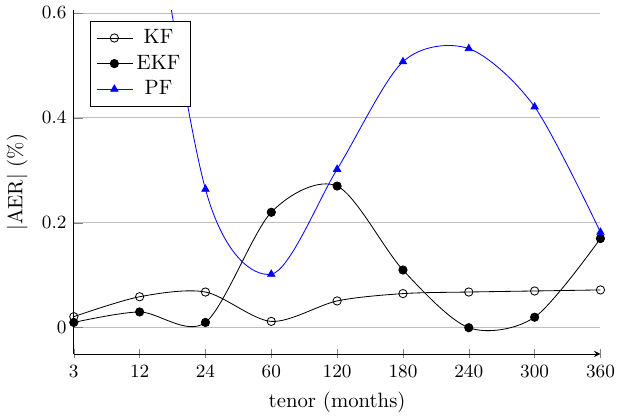} %
    \label{subfig:panel4}
   \end{subfigure}
  \label{fig-aer-4}
\end{figure}

        In Figure \ref{fig6_5_1}, we show the average excess return (AER) obtained from the evolution of forward rate curves that indicates the excess rate of the bond prices over the risk-free prices. The AER theoretically improves the soundness of the model and minimizes arbitrage opportunities in the dynamics of forward  rate curves.  The value of the AER shown in Figure  \ref{fig6_5_1} is obtained from the trained model with arbitrage-free regularization ($\lambda = 1$). The trained models without arbitrage regularization ($\lambda = 0$) have very high AER values which are not comparable. Among the three models, the AER for the KF model is significantly lower than the AER for the EKF and PF, particularly on the training set. We use panel-specific y-axis scales in Figure 5.4 to highlight the shape of each model's loss across maturities. The KF model has relatively stable AER across maturities in both the training and test sets, while the EKF and PF vary over maturities. In absolute terms, the magnitude of the AER is highest for the EKF at longer maturities, whereas the PF exhibits the largest AER at short maturities. From the perspective of models, the consistency and the minimum value of AER provided by the Kalman filter across the training result and the testing result indicates that forecasting in the yield-space with arbitrage-free regularization is more robust than the nonlinear filter models forecasting in the price-space.  This is consistent with the stronger stability of arbitrage-regularized dynamics when forecasting in yield-space under the linear state--space representation.

 Sensitivity to the arbitrage penalty $\lambda$, error-model choices (Gaussian vs.\ MGGD), and PF settings is reported in Section~\ref{sec:robustness}.  Therefore, we considered only the cases $\lambda=0$ (AER off) and $\lambda=1$ (AER on). We next consider the prediction performance of the models applied to corporate bond data similarly.

 	\subsection{Main Results (Corporates)}
 	We next apply our models to the credit spreads over treasuries for 10 corporate issuers and examine forecasting performance. Table \ref{Table9} shows the $5$-day-ahead forecasting results on data from ten corporate bond issuers with the predicted corporate spread calculated by subtracting the predicted Treasury yield from the predicted corporate yield and comparing to the observed value. For the forecasting results of corporate data, we show the predicted spread errors (predicted corporate yields $-$ predicted Treasury yields) are less than $14$ bps in $5$-day-ahead forecasting. Since credit risk factors are not included, and the corporate data contains only around 10 to 30 daily bonds, the forecasting performance of corporate data is not comparable to that of Treasury data. In the case of corporates the yield data contains seems to contain more information than the corporate bond prices and we find that the KF model  significantly outperforms the EKF and PF models.
 	
 	In related work, other models incorporating credit risk factors such as \citet{duffie2005credit} show prediction errors around $100$ basis points on short-term corporate bond yields and around $9$ basis points on long-term. \citet{duffee1999estimating} investigates 161 firm's bonds on monthly basis and shows the RMSE forecasting yield error in $34.56$ bps for 6-month maturity and $7.77$ bps for 30-year maturities, using the Kalman filter and a CIR model as interest rate in $1$-month-ahead forecasting.  However, \citet{duffee1999estimating} does not provide the out-of-sample tests.  \citet{ganguli2017machine} study corporate-bond forecasting with a 61-feature trade-level dataset and report results using a weighted error metric (WEPS); because the metric and features differ from ours, we do not compare levels.  As noted in \citet{diebold2006forecasting}, there is a persistent discrepancy between actual bond prices and the prices estimated from term structure models for the Treasury bonds. We do not smooth the observed prices so the discrepancy in the corporate bonds would be much larger than the Treasury bonds due to credit risk and/or liquidity problem.

 	\begin{table}[ht]
 		\centering
 		\caption{Testing result of 5-day-ahead forecasting: spread error (bps) and price error (dollar)}\label{Table9}
 		\renewcommand\arraystretch{0.8}
 		\begin{adjustbox}{max width=\linewidth,center}
 			\begin{tabular}{cccccccccccccc}
 				\hline
 				Ticker&\multicolumn{2}{|c|}{MAPE}&\multicolumn{2}{c|}{STDV}&\multicolumn{1}{c|}{MPE}&\multicolumn{1}{c}{hit rate}&\multicolumn{1}{|c}{Ticker}&\multicolumn{2}{|c|}{MAPE}&\multicolumn{2}{c|}{STDV}&\multicolumn{1}{c|}{MPE}&\multicolumn{1}{c}{hit rate}\\
 				\hline
 				Model&\multicolumn{1}{|c}{Spread}&\multicolumn{1}{c|}{Price}&\multicolumn{1}{c}{Spread}&\multicolumn{1}{c|}{Price}&\multicolumn{1}{c|}{$\%$}&\multicolumn{1}{c|}{ $\leq \$ 0.25$}&\multicolumn{1}{c}{Model}&\multicolumn{1}{|c}{Spread}&\multicolumn{1}{c|}{Price}&\multicolumn{1}{c}{Spread}&\multicolumn{1}{c|}{Price}&\multicolumn{1}{c}{$\%$}&\multicolumn{1}{|c}{ $\leq \$ 0.25$}\\
 				\hline
 				AAPL &  &  &  &  &  &  & AAPL &  &  &  &  &  & \\
 				KF ($\lambda$=0) & 5.27 & 0.241 & 2.54 & 0.353 & 0.231 & 68.1$\%$ & KF ($\lambda$=1) & 6.7 & 0.246 & 3.02 & 0.359 & 0.348 & 59.7$\%$ \\
 				EKF ($\lambda$=0) & 11.47 & 0.397 & 5.08 & 0.635 & 0.379 & 52.9$\%$ & EKF ($\lambda$=1) & 10.52 & 0.355 & 3.51 & 0.523 & 0.341 & 53.0$\%$ \\
 				PF ($\lambda$=0) & 10.4 & 0.259 & 3.44 & 0.386 & 0.248 & 65.8$\%$ & PF ($\lambda$=1) & 9.44 & 0.351 & 2.58 & 0.533 & 0.288 & 60.6$\%$ \\
 				C &  &  &  &  &  &  & C &  &  &  &  &  &  \\
 				KF ($\lambda$=0) & 9.93 & 0.447 & 2.92 & 0.713 & 0.404 & 45.5$\%$ & KF ($\lambda$=1) & 9.34 & 0.449 & 1.92 & 0.704 & 0.390 & 46.0$\%$ \\
 				EKF ($\lambda$=0) & 10.66 & 0.454 & 3.27 & 0.726 & 0.411 & 43.8$\%$ & EKF ($\lambda$=1) & 13 & 0.491 & 5.06 & 0.793 & 0.408 & 45.3$\%$ \\
 				PF ($\lambda$=0) & 11.05 & 0.49 & 2.44 & 0.816 & 0.440 & 45.1$\%$ & PF ($\lambda$=1) & 14.03 & 0.534 & 6.45 & 0.898 & 0.412 & 44.1$\%$ \\
 				DIS &  &  &  &  &  &  & DIS &  &  &  &  &  &  \\
 				KF ($\lambda$=0) & 7.83 & 0.365 & 3.46 & 0.847 & 0.341 & 62.0$\%$ & KF ($\lambda$=1) & 11.76 & 0.369 & 3.48 & 0.587 & 0.334 & 62.5$\%$ \\
 				EKF ($\lambda$=0) & 11.24 & 0.372 & 4.81 & 0.67 & 0.351 & 53.2$\%$ & EKF ($\lambda$=1) & 11.02 & 0.394 & 3.94 & 0.757 & 0.352 & 55.2$\%$ \\
 				PF ($\lambda$=0) & 11.33 & 0.383 & 4.44 & 0.721 & 0.358 & 54.7$\%$ & PF ($\lambda$=1) & 9.61 & 0.382 & 3.52 & 0.812 & 0.363 & 54.0$\%$ \\
 				GS &  &  &  &  &  &  & GS &  &  &  &  &  &  \\
 				KF ($\lambda$=0) & 8.75 & 0.426 & 2.89 & 0.615 & 0.388 & 47.2$\%$ & KF ($\lambda$=1) & 8.66 & 0.428 & 1.71 & 0.604 & 0.402 & 45.6$\%$ \\
 				EKF ($\lambda$=0) & 9.16 & 0.437 & 1.96 & 0.649 & 0.402 & 47.3$\%$ & EKF ($\lambda$=1) & 10.81 & 0.434 & 2.72 & 0.658 & 0.402 & 48.0$\%$ \\
 				PF ($\lambda$=0) & 10.24 & 0.433 & 3.06 & 0.653 & 0.397 & 47.7$\%$ & PF ($\lambda$=1) & 11.03 & 0.475 & 2.94 & 0.727 & 0.400 & 47.1$\%$ \\
 				JNJ &  &  &  &  &  &  & JNJ &  &  &  &  &  &  \\
 				KF ($\lambda$=0) & 7.38 & 0.454 & 3.78 & 0.706 & 0.412 & 47.0$\%$ & KF ($\lambda$=1) & 8.06 & 0.429 & 3.46 & 0.645 & 0.400 & 47.4$\%$ \\
 				EKF ($\lambda$=0) & 10.61 & 0.54 & 6.13 & 0.879 & 0.496 & 40.4$\%$ & EKF ($\lambda$=1) & 10.21 & 0.541 & 3.58 & 0.882 & 0.518 & 42.0$\%$ \\
 				PF ($\lambda$=0) & 9.96 & 0.578 & 4.01 & 0.937 & 0.527 & 41.6$\%$ & PF ($\lambda$=1) & 11.01 & 0.607 & 3.9 & 1.012 & 0.475 & 41.9$\%$ \\
 				JPM &  &  &  &  &  &  & JPM &  &  &  &  &  &  \\
 				KF ($\lambda$=0) & 6.46 & 0.346 & 1.77 & 0.616 & 0.307 & 58.7$\%$ & KF ($\lambda$=1) & 8.49 & 0.45 & 4.12 & 0.914 & 0.324 & 55.9$\%$ \\
 				EKF ($\lambda$=0) & 10.07 & 0.473 & 3.95 & 0.909 & 0.412 & 52.1$\%$ & EKF ($\lambda$=1) & 11.31 & 0.508 & 3.38 & 0.96 & 0.398 & 48.3$\%$ \\
 				PF ($\lambda$=0) & 10.62 & 0.491 & 3.58 & 0.914 & 0.429 & 50.9$\%$ & PF ($\lambda$=1) & 12.63 & 0.482 & 4.97 & 0.926 & 0.430 & 50.0$\%$ \\
 				MSFT &  &  &  &  &  &  & MSFT &  &  &  &  &  &  \\
 				KF ($\lambda$=0) & 5.62 & 0.343 & 2.77 & 0.492 & 0.325 & 52.5$\%$ & KF ($\lambda$=1) & 8.34 & 0.448 & 4.96 & 0.807 & 0.331 & 52.3$\%$ \\
 				EKF ($\lambda$=0) & 10.56 & 0.441 & 3.37 & 0.653 & 0.419 & 45.9$\%$ & EKF ($\lambda$=1) & 9.9 & 0.429 & 2.84 & 0.662 & 0.434 & 43.4$\%$ \\
 				PF ($\lambda$=0) & 11.6 & 0.393 & 4.14 & 0.594 & 0.373 & 50.1$\%$ & PF ($\lambda$=1) & 10.84 & 0.433 & 3.6 & 0.67 & 0.406 & 47.4$\%$ \\
 				T &  &  &  &  &  &  & T &  &  &  &  &  &  \\
 				KF ($\lambda$=0) & 9.41 & 0.45 & 5.62 & 1.022 & 0.398 & 57.1$\%$ & KF ($\lambda$=1) & 10.26 & 0.407 & 5.11 & 0.85 & 0.370 & 59.6$\%$ \\
 				EKF ($\lambda$=0) & 10.56 & 0.489 & 4.59 & 0.951 & 0.440 & 49.8$\%$ & EKF ($\lambda$=1) & 14.14 & 0.611 & 5 & 1.057 & 0.431 & 48.7$\%$ \\
 				PF ($\lambda$=0) & 12.81 & 0.389 & 5.31 & 0.75 & 0.353 & 57.5$\%$ & PF ($\lambda$=1) & 14.04 & 0.53 & 6.55 & 1.032 & 0.361 & 57.5$\%$ \\
 				UNH &  &  &  &  &  &  & UNH &  &  &  &  &  &  \\
 				KF ($\lambda$=0) & 8.34 & 0.404 & 3.63 & 0.881 & 0.377 & 54.2$\%$ & KF ($\lambda$=1) & 10.15 & 0.433 & 5.18 & 0.909 & 0.343 & 55.5$\%$ \\
 				EKF ($\lambda$=0) & 9.23 & 0.364 & 4.18 & 0.588 & 0.344 & 53.4$\%$ & EKF ($\lambda$=1) & 11.68 & 0.376 & 4.65 & 0.615 & 0.361 & 53.1$\%$ \\
 				PF ($\lambda$=0) & 10.96 & 0.378 & 3.09 & 0.744 & 0.353 & 56.5$\%$ & PF ($\lambda$=1) & 10.34 & 0.371 & 3.89 & 0.602 & 0.349 & 57.0$\%$ \\
 				WFC &  &  &  &  &  &  & WFC &  &  &  &  &  &  \\
 				KF ($\lambda$=0) & 7.03 & 0.368 & 2.82 & 0.778 & 0.330 & 68.0$\%$ & KF ($\lambda$=1) & 10.64 & 0.442 & 3.72 & 0.878 & 0.337 & 66.5$\%$ \\
 				EKF ($\lambda$=0) & 12.14 & 0.56 & 3.38 & 1.117 & 0.501 & 55.8$\%$ & EKF ($\lambda$=1) & 13.25 & 0.562 & 2.8 & 1.166 & 0.592 & 53.8$\%$ \\
 				PF ($\lambda$=0) & 12.77 & 0.538 & 3.34 & 1.029 & 0.487 & 54.5$\%$ & PF ($\lambda$=1) & 13.71 & 0.583 & 3.52 & 1.12 & 0.560 & 50.2$\%$ \\
 				\hline
 			\end{tabular}
 		\end{adjustbox}
 	\end{table}

	 \begin{figure}[ht]
 		\centering
 		\caption{Training result of U.S. Treasuries: Model loss}
 		\hspace*{-0cm}
 		\includegraphics[width=0.85\textwidth]{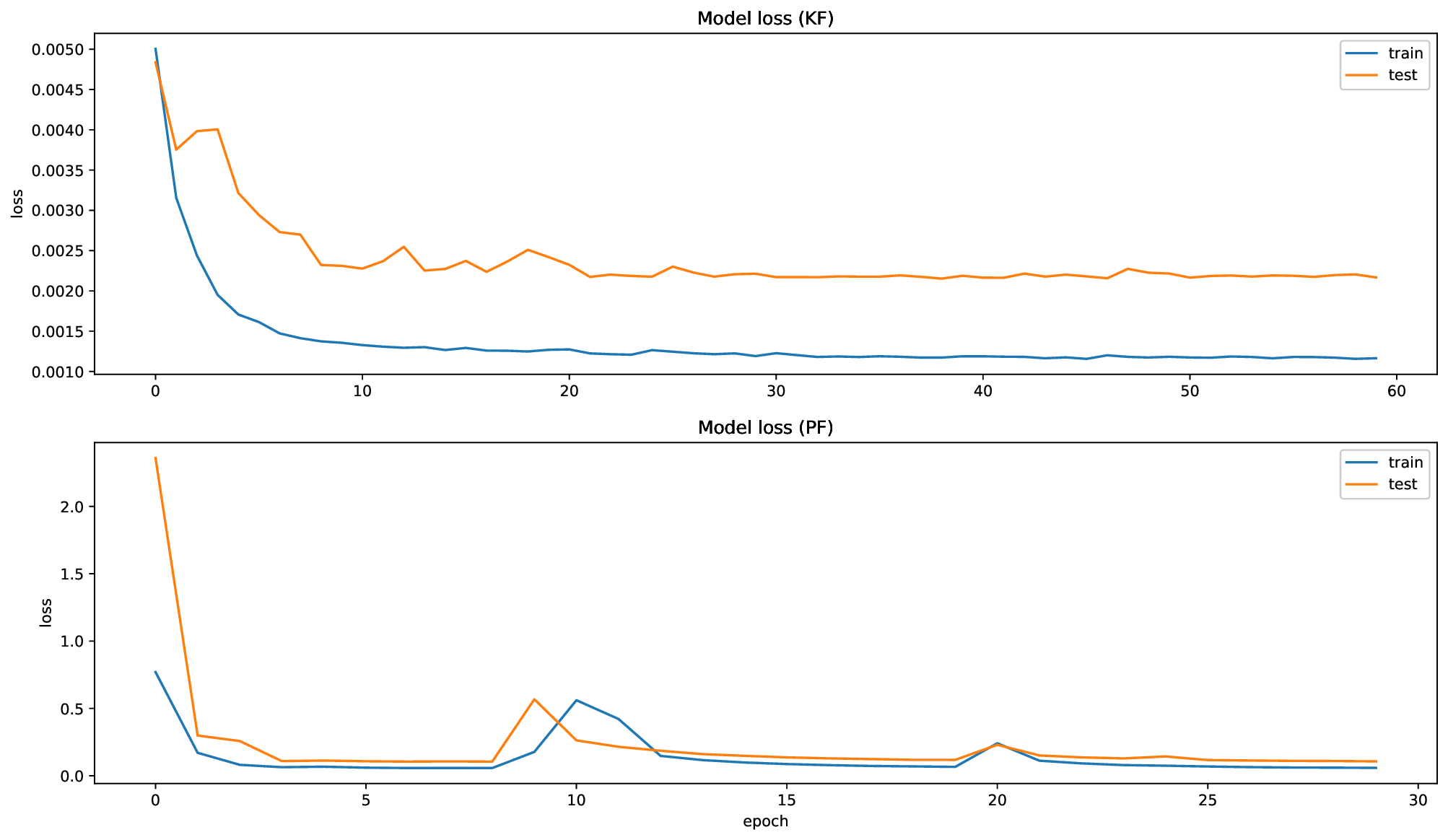}
 		\label{fig6_7_1} %
 	 \end{figure}

      \subsection{Robustness and Sensitivity}\label{sec:robustness}

      \paragraph{Training dynamics.}
         Figure~\ref{fig6_7_1} reports training and test losses.  The loss curves are stable with early stopping and show no overfitting for the horizons considered.

\paragraph{Arbitrage-penalty sensitivity.}
	We show the forecasting results of the bond prices using the model with Kalman filter by varying the value of the penalty ($\lambda$) in Table \ref{Table0_1} where we train the models to reach almost the same MSE and compare the hit rate. The MAE shows the mean absolute forecasting error of bond prices in dollars. We can see that the arbitrage-free regularization significantly improves the forecasting performance in $5$-day-ahead forecasting. The arbitrage-free penalty with $\lambda =1$ shows the best training MSE with overall the best hit rates. Increasing the value of $\lambda$ leads to increasing computational cost since the regularization term will dominate the target function and the training process takes longer to decrease the loss function.  
 	\begin{table}[ht]
 		\centering
 		\caption{Sensitivity to Arbitrage Penalty: U.S. Treasuries KF-Based Model (test set)}\label{Table0_1}
 		\begin{adjustbox}{max width=\linewidth, center}
 			\begin{tabular}{ccccccccccc}
 				\hline
 				\multicolumn{1}{c} {penalty} & \multicolumn{2}{c} {MAE} & \multicolumn{2}{c} {MSE} & \multicolumn{2}{c} {hit rate (\$0.10)} & \multicolumn{2}{c} {hit rate (\$0.25)} & \multicolumn{2}{c} {hit rate (\$0.50)} \\
 				\cline{2-11}
 				\multicolumn{1}{c} {$(\lambda)$} & \multicolumn{1}{c} {1-day} & \multicolumn{1}{c} {5-day} & \multicolumn{1}{c} {1-day} & \multicolumn{1}{c} {5-day} & \multicolumn{1}{c} {1-day} & \multicolumn{1}{c} {5-day} & \multicolumn{1}{c} {1-day} & \multicolumn{1}{c} {5-day}& \multicolumn{1}{c} {1-day} & \multicolumn{1}{c} {5-day}\\
 				\hline
 				0 & 0.1717 & 0.4428 & 0.1067 & 0.8084 & 54.46$\%$ & 32.66$\%$ & 83.40$\%$ & 58.25$\%$ & 93.43$\%$ & 75.32$\%$ \\
 				0.01 & 0.1709 & 0.4244 & 0.1043 & 0.9193 & 54.68$\%$ & 34.85$\%$ & 83.18$\%$ & 61.37$\%$ & 93.58$\%$ & 77.95$\%$ \\
 				0.1 & 0.1712 & 0.4079 & 0.1054 & 0.8071 & 54.91$\%$ & 37.16$\%$ & 83.11$\%$ & 63.21$\%$ & 93.50$\%$ & 80.08$\%$ \\
 				0.5 & 0.1960 & 0.4429 & 0.1859 & 1.2945 & 54.02$\%$ & 39.55$\%$ & 82.99$\%$ & 65.26$\%$ & 92.11$\%$ & 79.97$\%$ \\
 				1 & 0.1750 & 0.3630 & 0.1136 & 0.6886 & 54.43$\%$ & 40.30$\%$ & 83.40$\%$ & 66.77$\%$ & 93.03$\%$ & 83.14$\%$ \\
 				1.5 & 0.2119 & 0.4392 & 0.2724 & 1.4783 & 54.49$\%$ & 39.46$\%$ & 82.46$\%$ & 65.79$\%$ & 91.89$\%$ & 81.25$\%$ \\
 				10 & 0.1956 & 0.4173 & 0.1750 & 1.1795 & 52.87$\%$ & 40.94$\%$ & 81.54$\%$ & 67.44$\%$ & 92.55$\%$ & 82.36$\%$ \\
 				\hline
 			\end{tabular}
 		\end{adjustbox}
 	\end{table}

\begin{figure}[ht]
 		\centering
 		\caption{U.S. Treasuries: path of state variables of 1-day-ahead forecasting}
 		\vspace*{-.0cm}
 		\includegraphics[width=0.9\linewidth, height=3.5in]{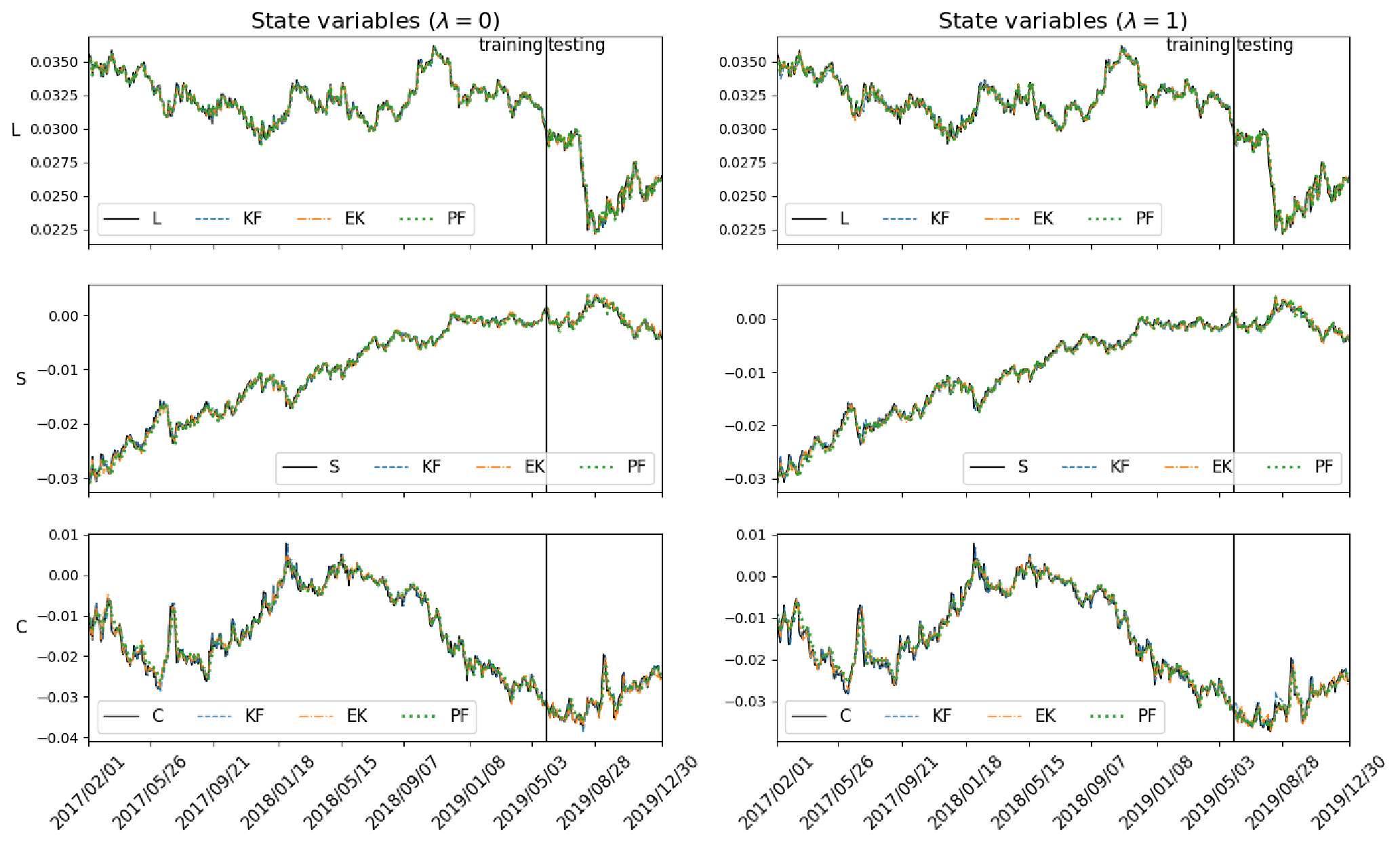}
 		\label{fig6_1_1} %
 	\end{figure}

\begin{figure}[ht]
 		\centering
 		\vspace*{-.0cm}
 		\caption{U.S. Treasuries: path of state variables of 5-day-ahead forecasting}
 		\vspace*{-.0cm}
 		\includegraphics[width=0.9\linewidth, height=3.5in]{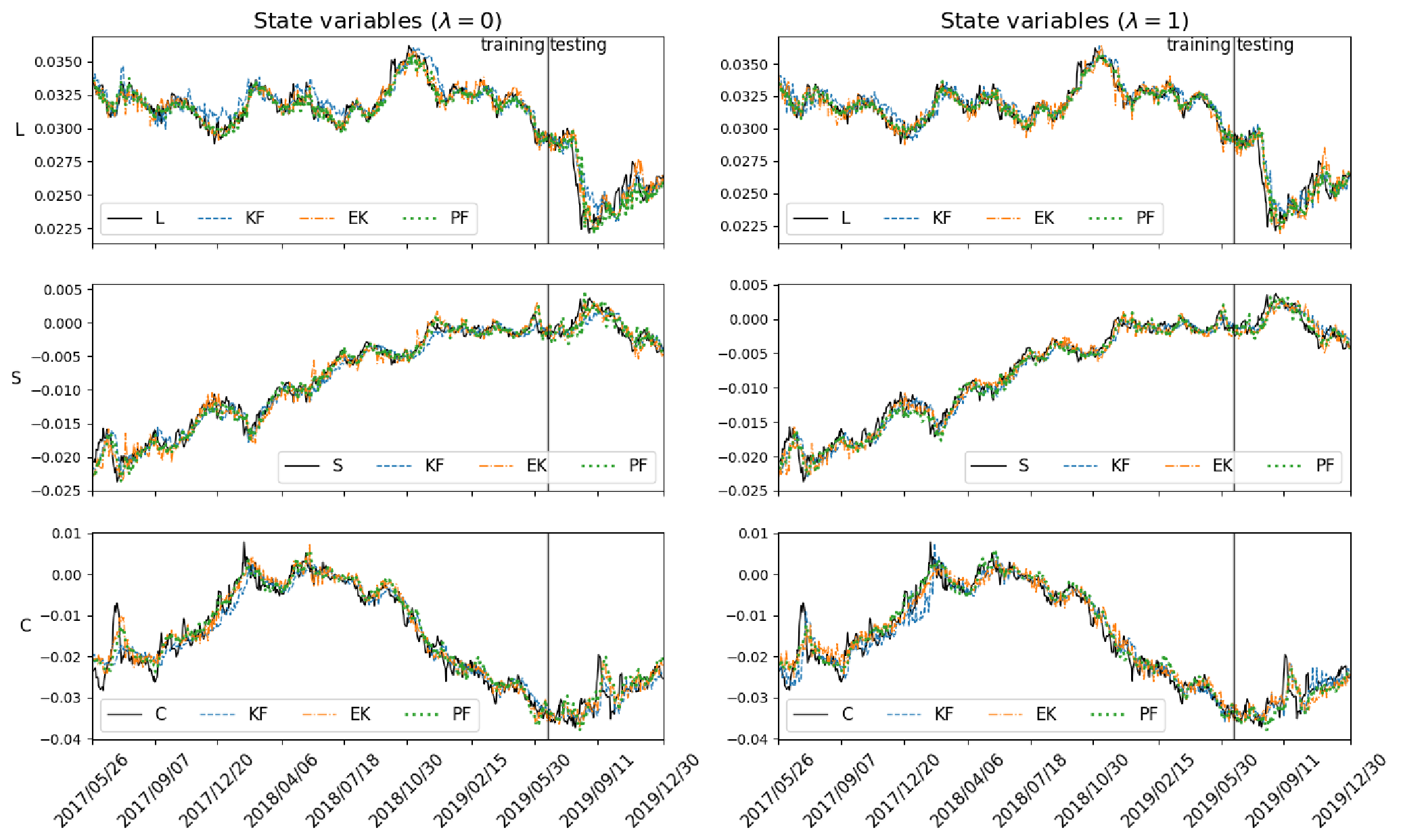}
 		\label{fig6_1_2} %
 	\end{figure}

	In Figure \ref{fig6_1_1} and \ref{fig6_1_2}, we show the $h$-day-ahead forecasting of state variables $X(t)=(X_1(t),X_2(t),X_3(t))^\top$ comparing to the observed state variables on a daily basis as short-term, mid-term and long-term levels. In Figure \ref{fig6_1_1} showing the 1-day-ahead forecasting, the difference between the forecasted result and observed results are undiscernible. However, in Figure \ref{fig6_1_2} showing the 5-day-ahead forecasting, we see that the forecasted result with arbitrage regularization (AR) is closer to the observed results. The forecasted paths of state variables for the EKF-based model shows more oscillation than the forecasted paths obtained using the PF-based model.

 	\begin{figure}[ht]
 		\centering
 		\vspace*{0.0cm}
 		\caption{U.S. Treasuries: yield curves of 1-day-ahead forecasting}
 		\includegraphics[width=0.9\linewidth, height=3.5in]{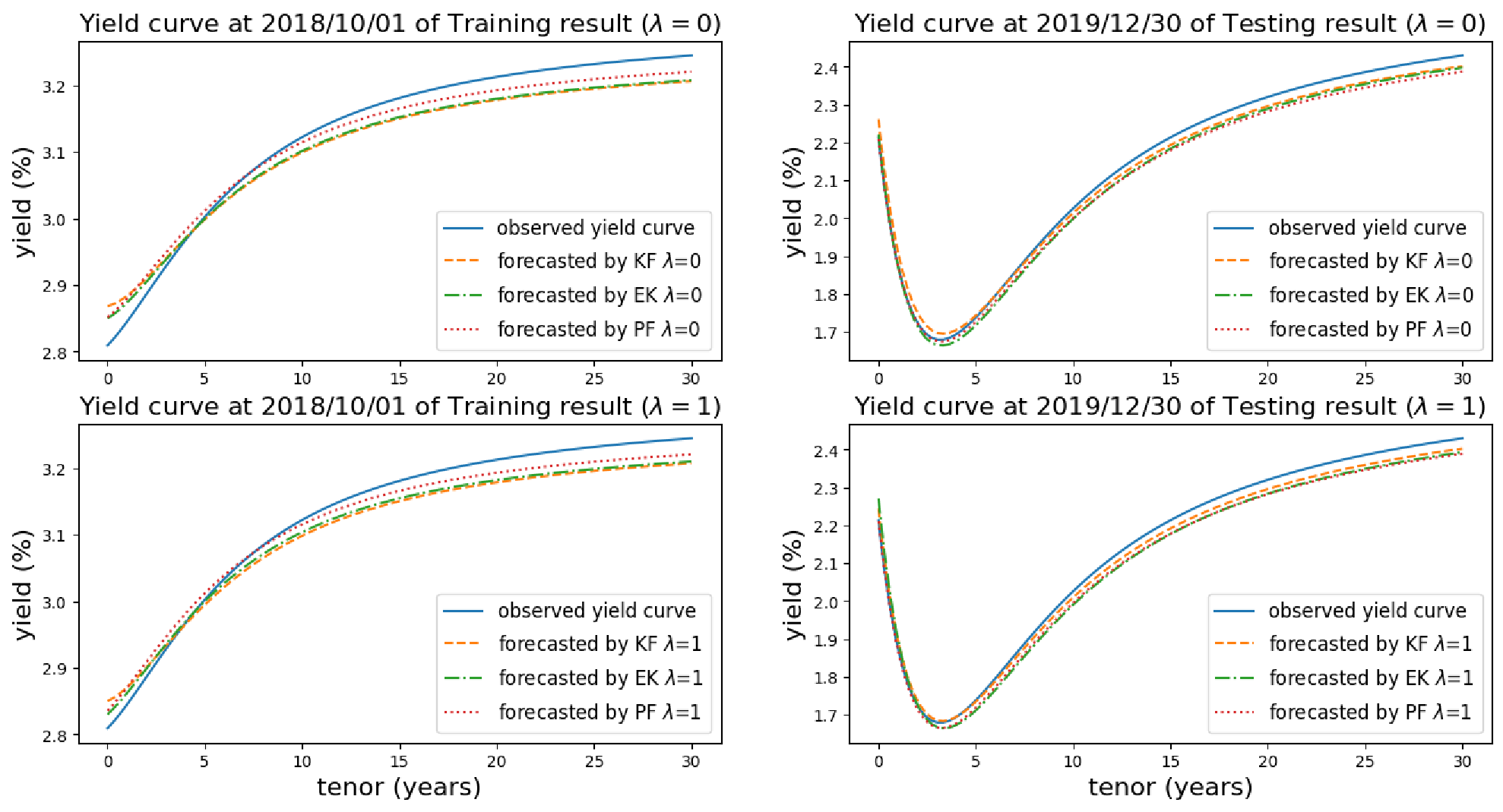}
 		\label{fig6_6} %
 	\end{figure}
 	
 	\begin{figure}[ht]
 		\centering
 		\vspace*{0.0cm}
 		\caption{U.S. Treasuries: yield curves of 5-day-ahead forecasting}
 		\includegraphics[width=0.9\linewidth, height=3.5in]{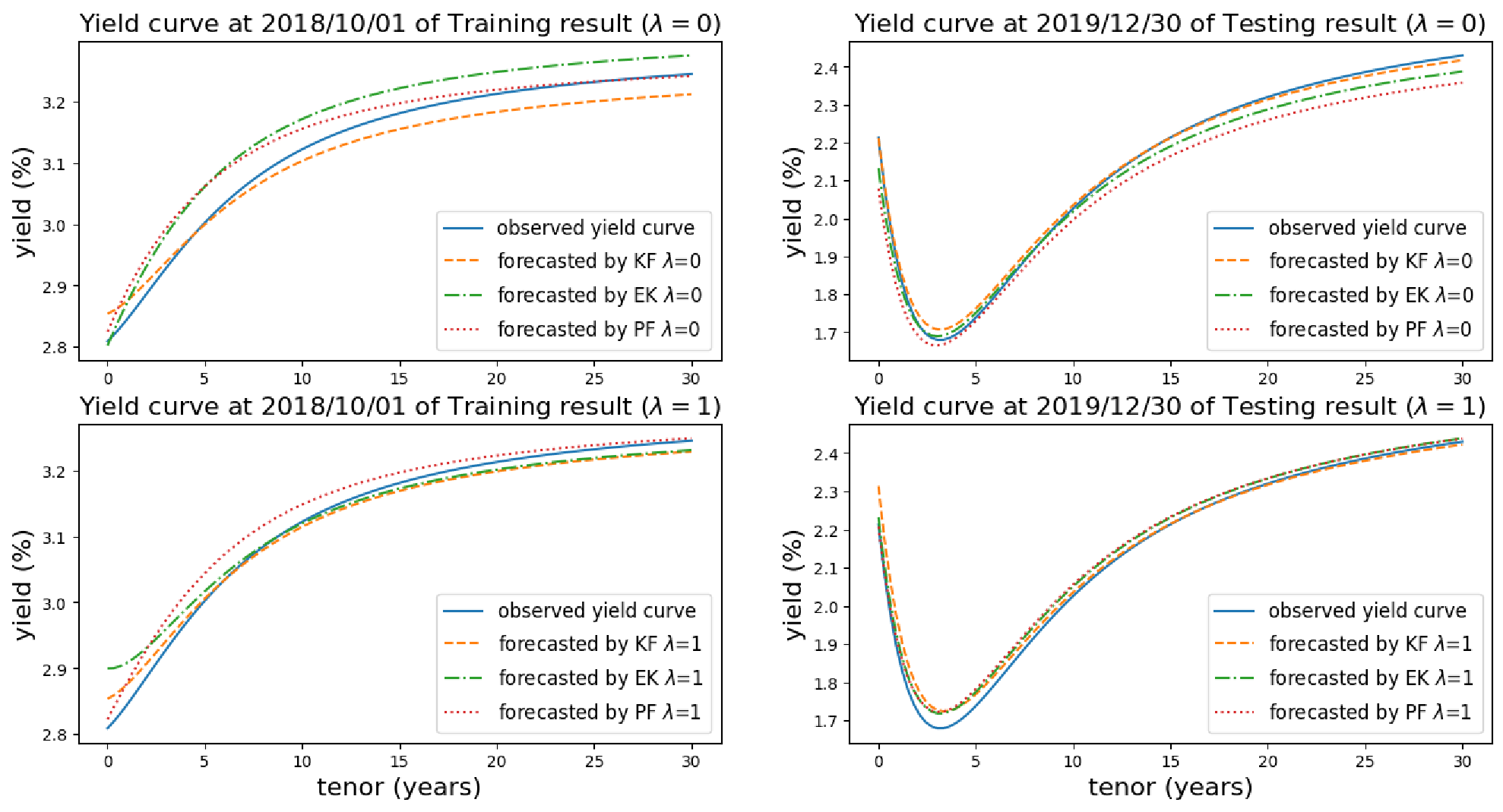}
 		\label{fig6_61} %
 	\end{figure}

	Figure \ref{fig6_6} and Figure \ref{fig6_61} present two examples of predicted yield curves: the left one is an increasing yield curve from the training set and the right one is an inverted humped yield curve from the testing set. From Figure \ref{fig6_6} and Figure \ref{fig6_61}, we can see that the forecasted yield curves with the restriction of arbitrage-free regularization show higher accuracy and this effect is more obvious in $5$-day-ahead forecasting.
 
\begin{figure}[ht]
 		\centering
 		\caption{U.S. Treasuries: state parameters (Kalman filter)}
 		\hspace*{-0cm}
 		\includegraphics[width=0.9\linewidth, height=3.5in]{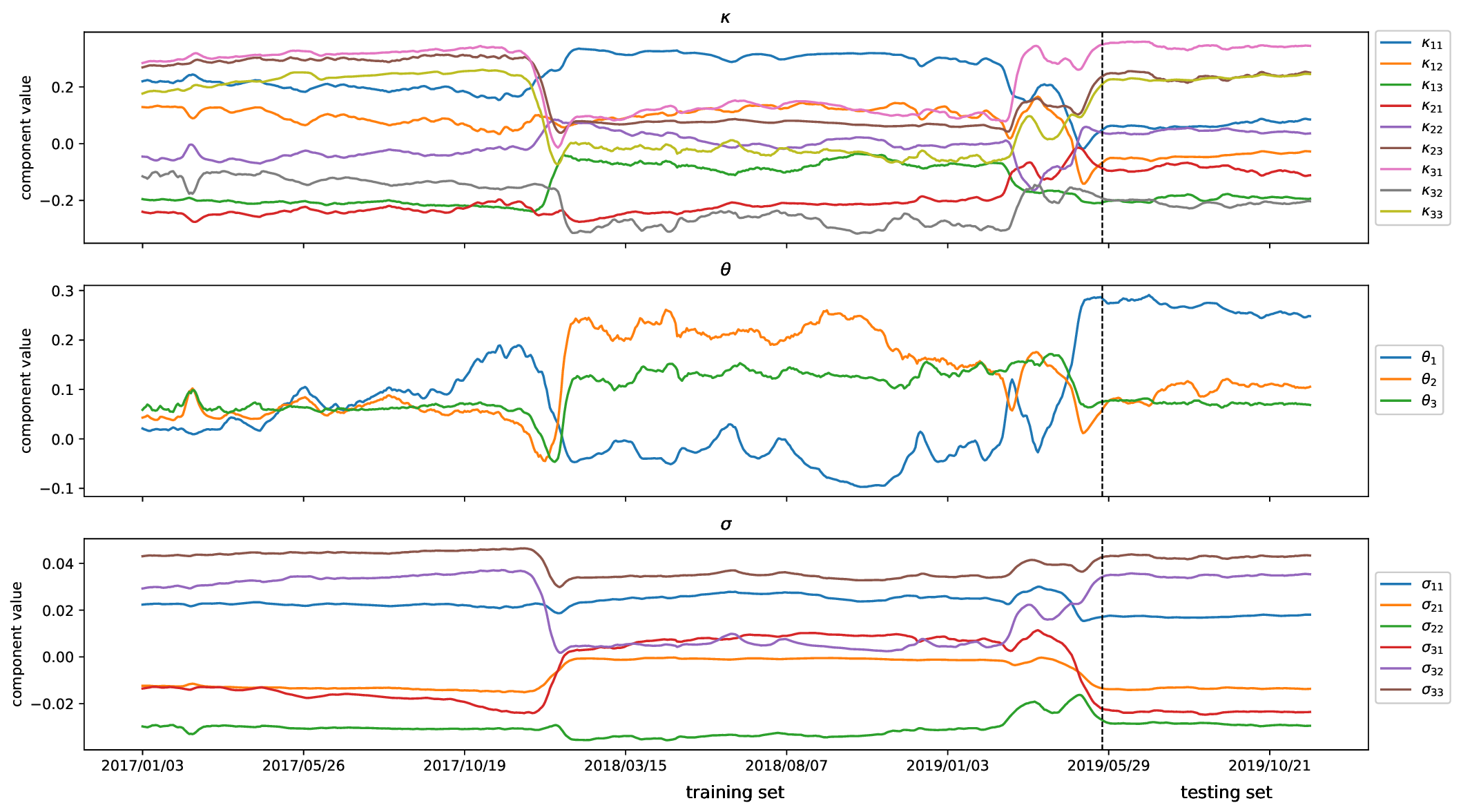}
 		\label{fig6_7_2} %
 	\end{figure}

\begin{figure}[ht]
 		\centering
 		\caption{U.S. Treasuries: State parameters (Kalman filter + arbitrage regularization)}
 		\hspace*{-0cm}
 		\includegraphics[width=0.9\linewidth, height=3.5in]{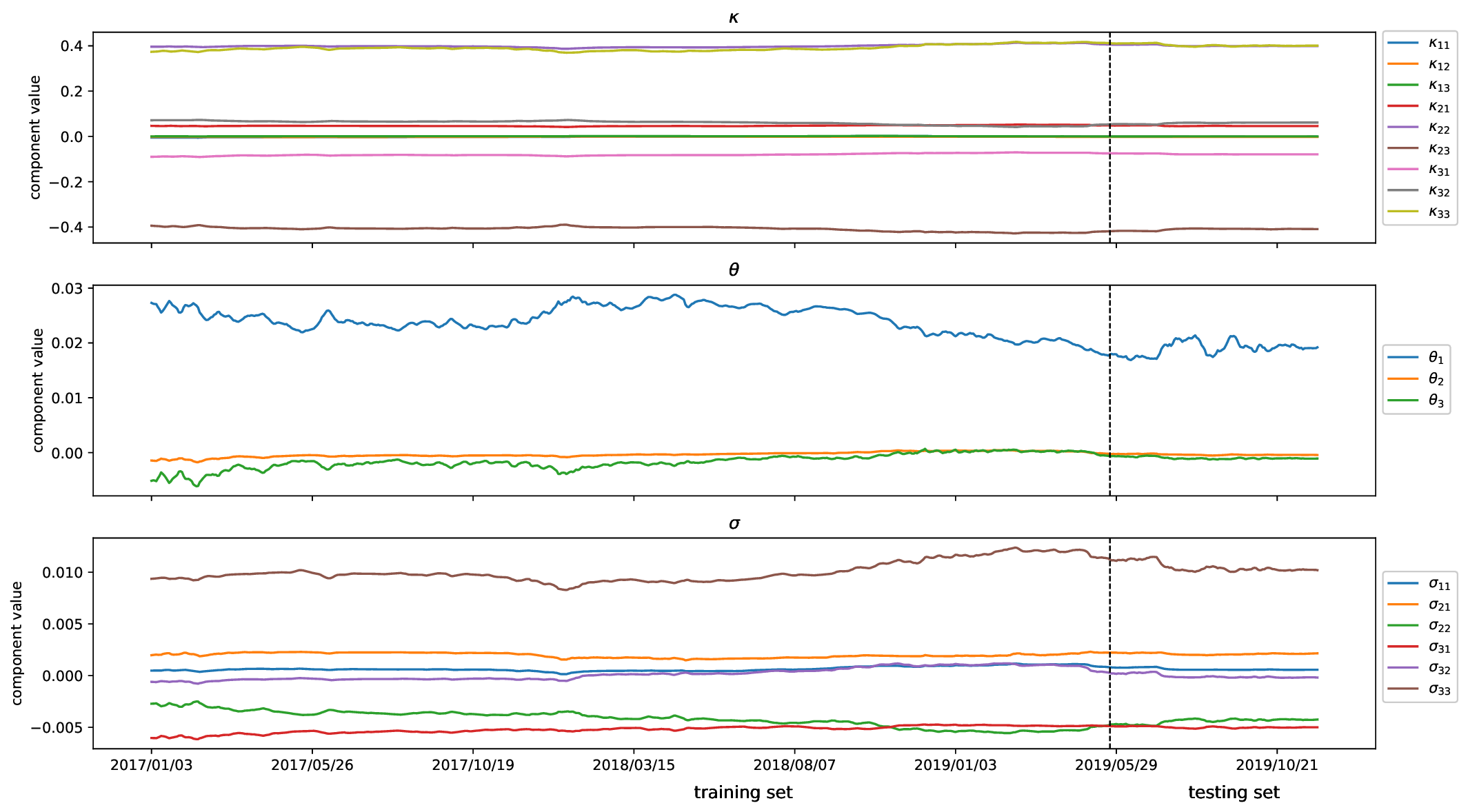}
 		\label{fig6_7_3} %
 	\end{figure}

 	  Figures \ref{fig6_7_2} and \ref{fig6_7_3} show the variation of the state parameters: $\kappa_t$, $\theta_t$ and $\sigma_t$ obtained from the yield prediction and bond price prediction models using the Kalman filter with and without arbitrage-free regularization.  Visual inspection suggests regime dependence without AER and stabilization with AER. A formal test (e.g., Markov-switching or Bai--Perron tests) is beyond scope of this paper but left for future work.

	\begin{figure}[ht]
 		\centering
 		\caption{U.S. Treasuries: Q-Q plot (price error) of 1-day-ahead forecasting}
 		\hspace*{-0cm}
 		\includegraphics[width=0.9\linewidth, height=3.5in]{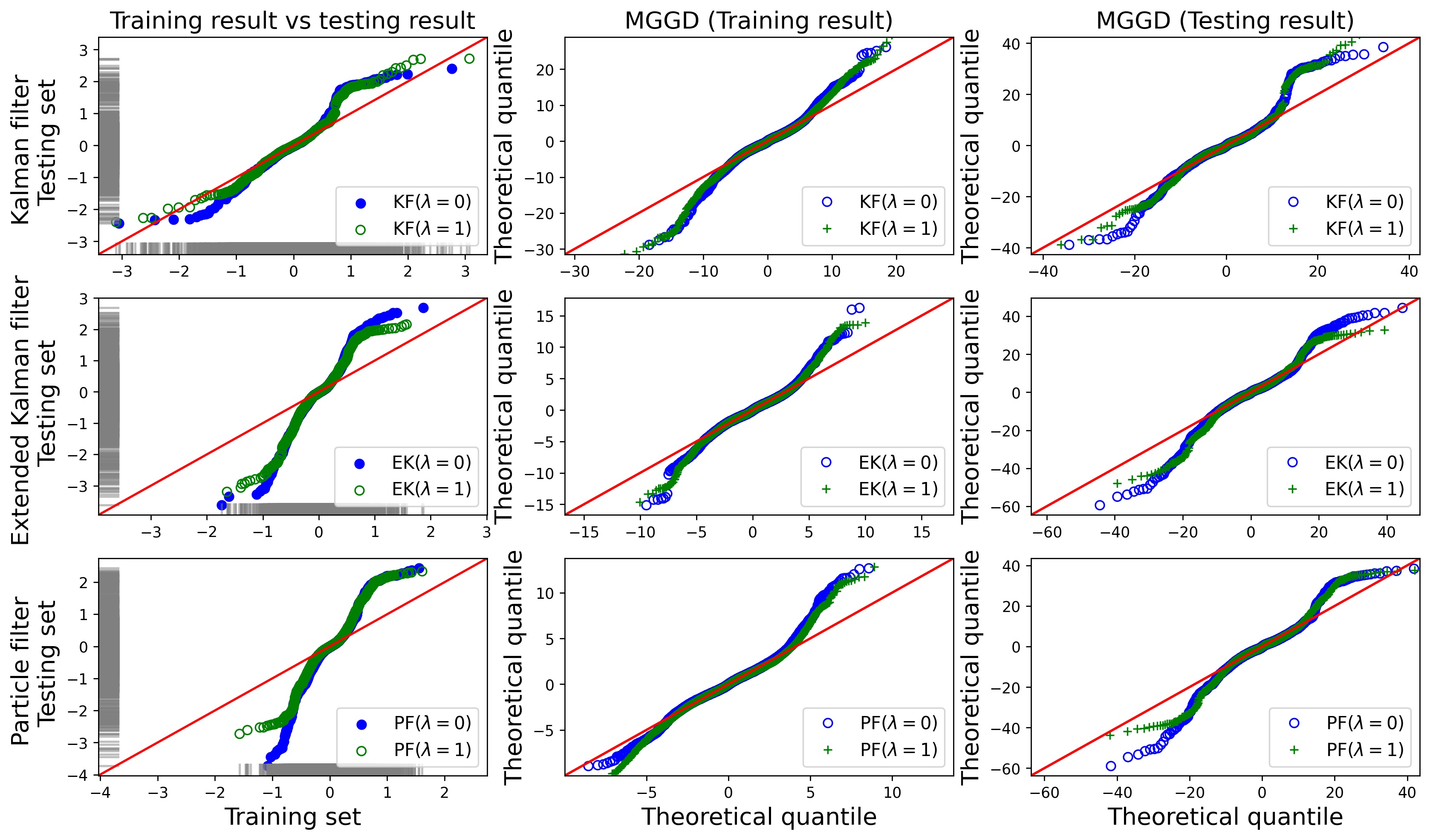}
 		\label{fig6_6_2}
 	\end{figure}
 	
 	\begin{figure}[ht]
 		\centering
 		\caption{U.S. Treasuries: yield error distribution of 1-day-ahead forecasting}
 		\hspace*{-0cm}
 		\includegraphics[width=0.9\linewidth, height=3.5in]{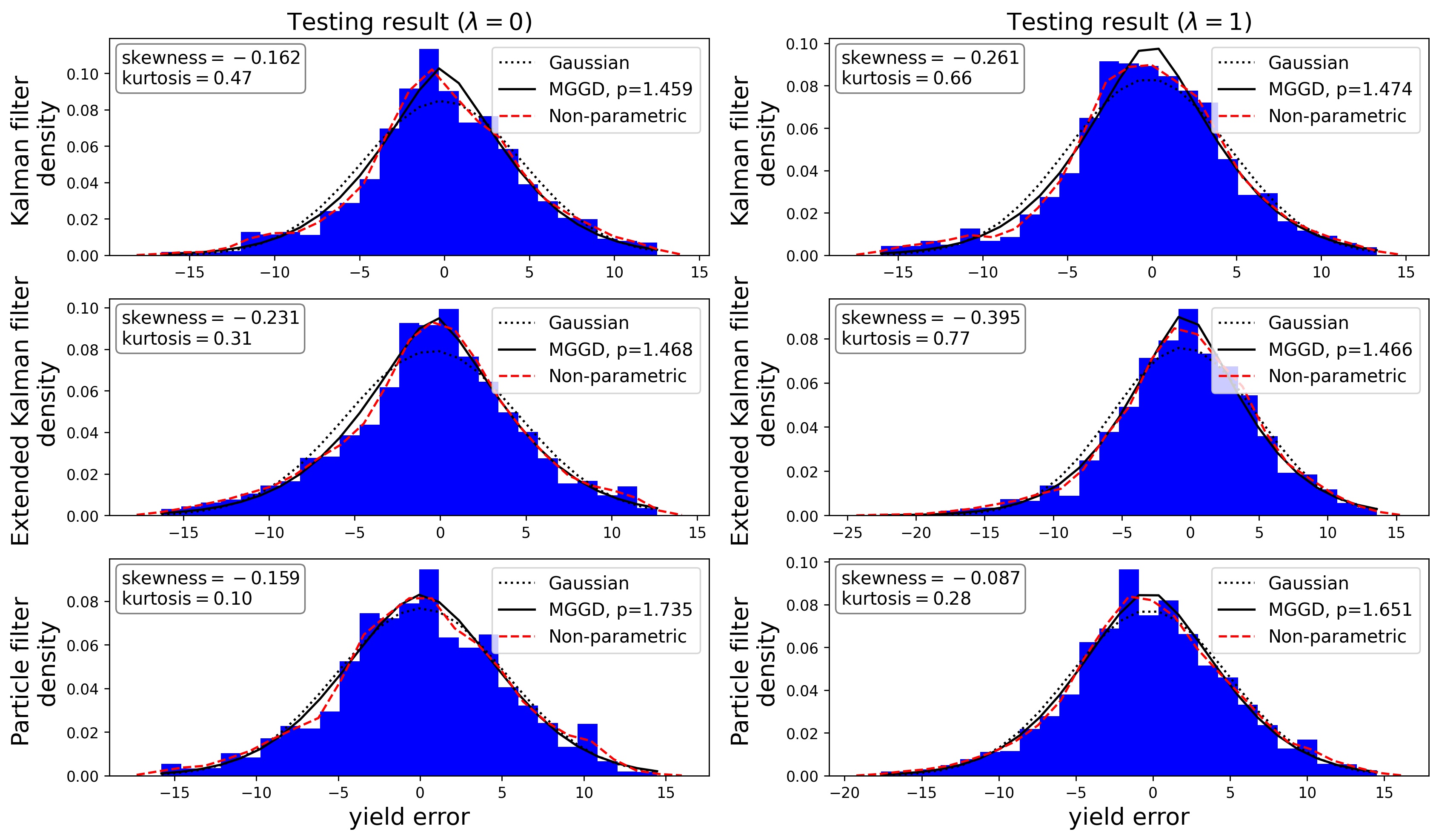}
 		\label{fig6_6_3} %
 	\end{figure}
 	
 	\begin{figure}[ht]
 		\centering
 		\caption{U.S. Treasuries: QQ-plot (yield error) of 1-day-ahead forecasting}
 		\hspace*{-0cm}
 		\includegraphics[width=0.9\linewidth, height=3.5in]{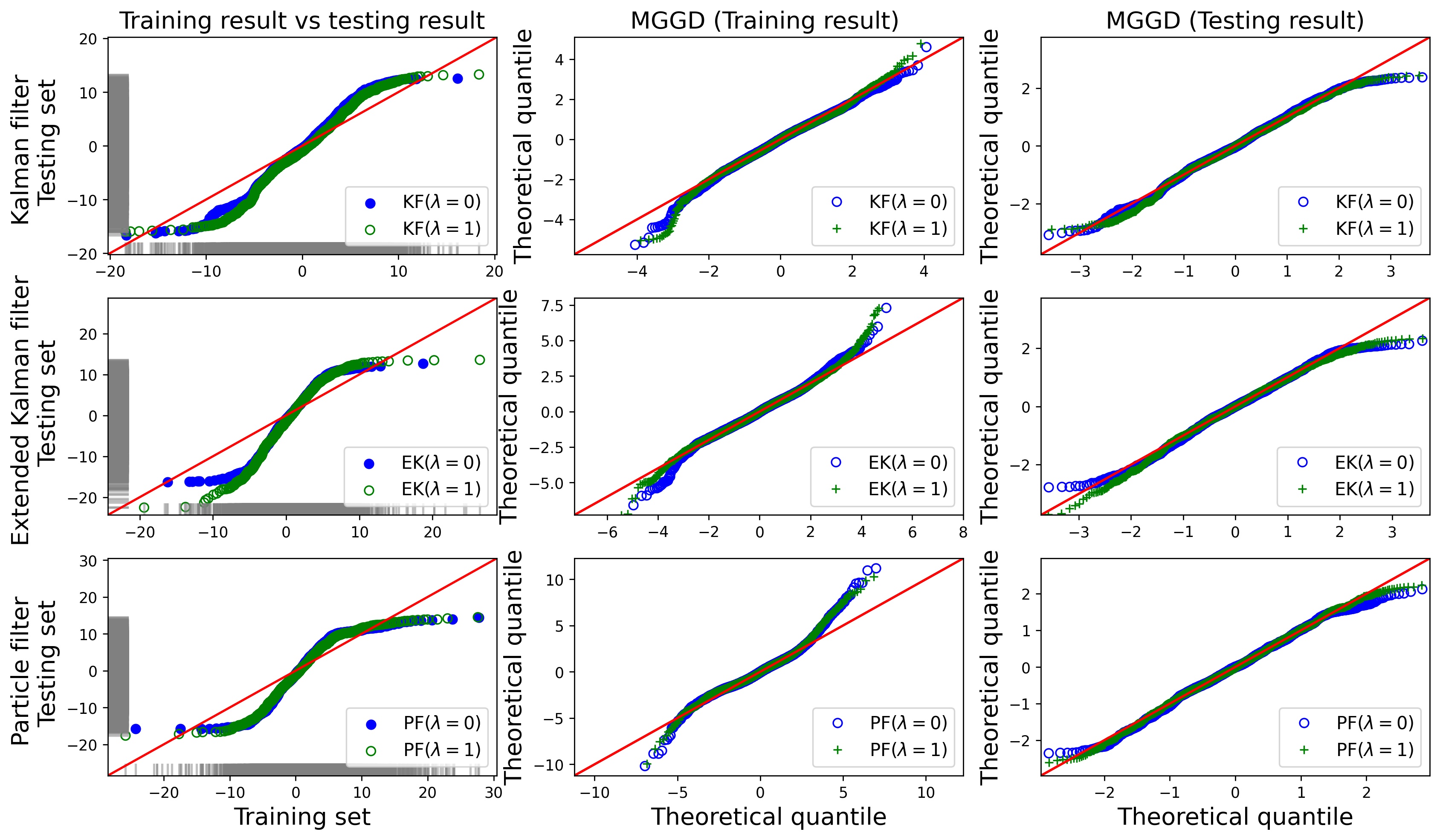}
 		\label{fig6_6_4}
 	\end{figure}

\textbf{Error model.} Results are qualitatively similar under Gaussian and MGGD error specifications; heavier tails tend to favor the PF at short maturities. Figures \ref{fig6_6_2}-\ref{fig6_6_4} and  \ref{fig6_6_1} fit the 1-day-ahead prediction errors of Treasury yields and bond prices with several candidate distributions and report the corresponding Q--Q plots. Figure  \ref{fig6_6_1} shows that bond-price prediction errors exhibit excess kurtosis, which is mitigated when the arbitrage penalty is active. Figure \ref{fig6_6_2} indicates that fat tails are pronounced for the nonlinear price-space models, while Figure \ref{fig6_6_3} shows that yield errors have relatively low kurtosis, again improved by arbitrage regularization. Comparing the PF results in Figures \ref{fig6_6_2} and \ref{fig6_6_1}, we find that the MGGD can accommodate both excess-kurtosis and low-kurtosis regimes, with the remaining challenge primarily related to tail thickness. For the excess-kurtosis behavior in the PF, a nonparametric observation model could be considered, while for fat tails more generally, jump-diffusion dynamics (as in \citet{brigo2007stochastic}) provide a natural extension.

\paragraph{Runtime and Implementation.} The run times for the three filter-assisted machine learning models are very different: the KF model runs in a couple of minutes, the EKF takes a few seconds to finish 1 epoch depending on the number of daily observations, and the training time of the PF increases exponentially as the number of particles increase which can take a few hours with 300 particles.\footnote{We run our models on Google Colab around 30$\sim$60 epochs which shows the optimal result without significant bias.}

\begin{figure}[ht]
 		\centering
 		\caption{U.S. Treasuries: Effective sample size with different regularization parameter $\lambda$}
 		\hspace*{-0cm} 		
 		\includegraphics[width=0.9\linewidth]{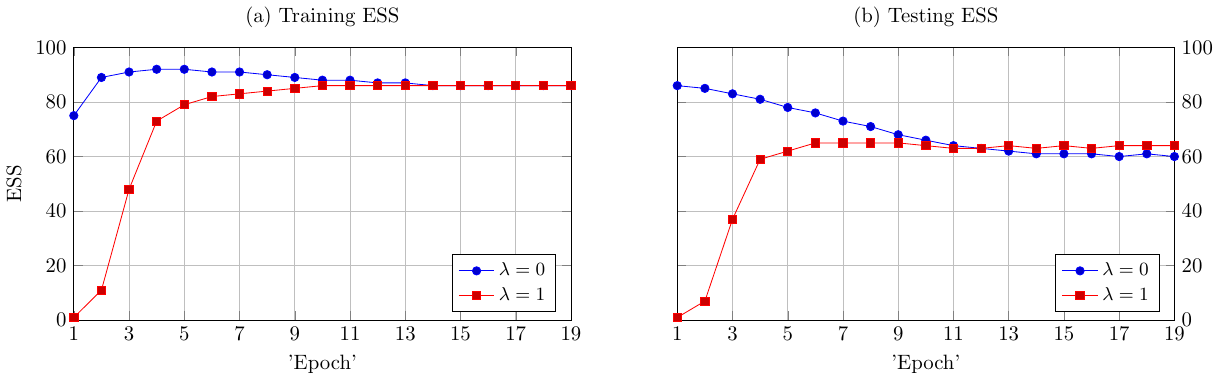}
 		\label{fig6_5_2}
 	\end{figure}

\begin{figure}[ht]
 		\centering
 		\caption{U.S. Treasuries: Effective sample size with different shape parameter $p$}
 		\hspace*{-0cm} 		
 		\includegraphics[width=0.9\linewidth]{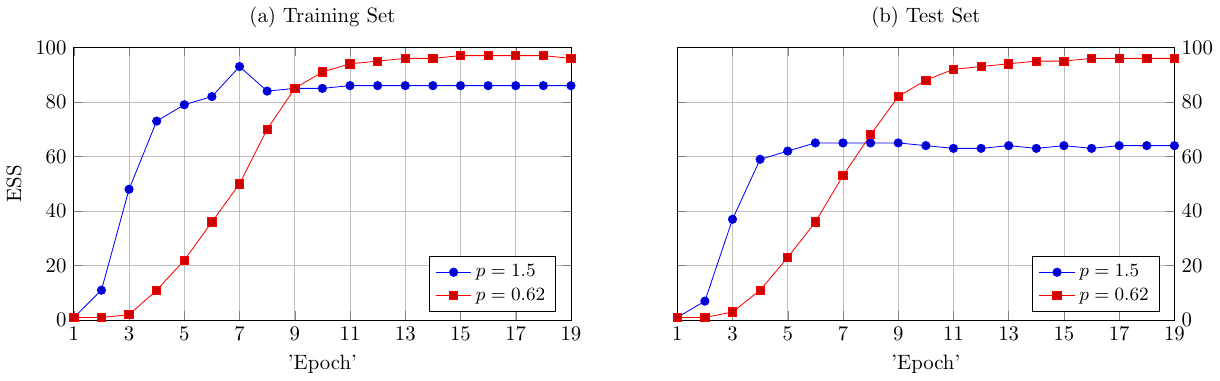}
 		\label{fig6_5_3}
 	\end{figure}

\begin{figure}[ht]
 		\centering
 		\caption{U.S.Treasuries: Price error distribution of 1-day-ahead forecasting}
 		\hspace*{-0cm}
 		\includegraphics[width=0.9\linewidth, height=3.5in]{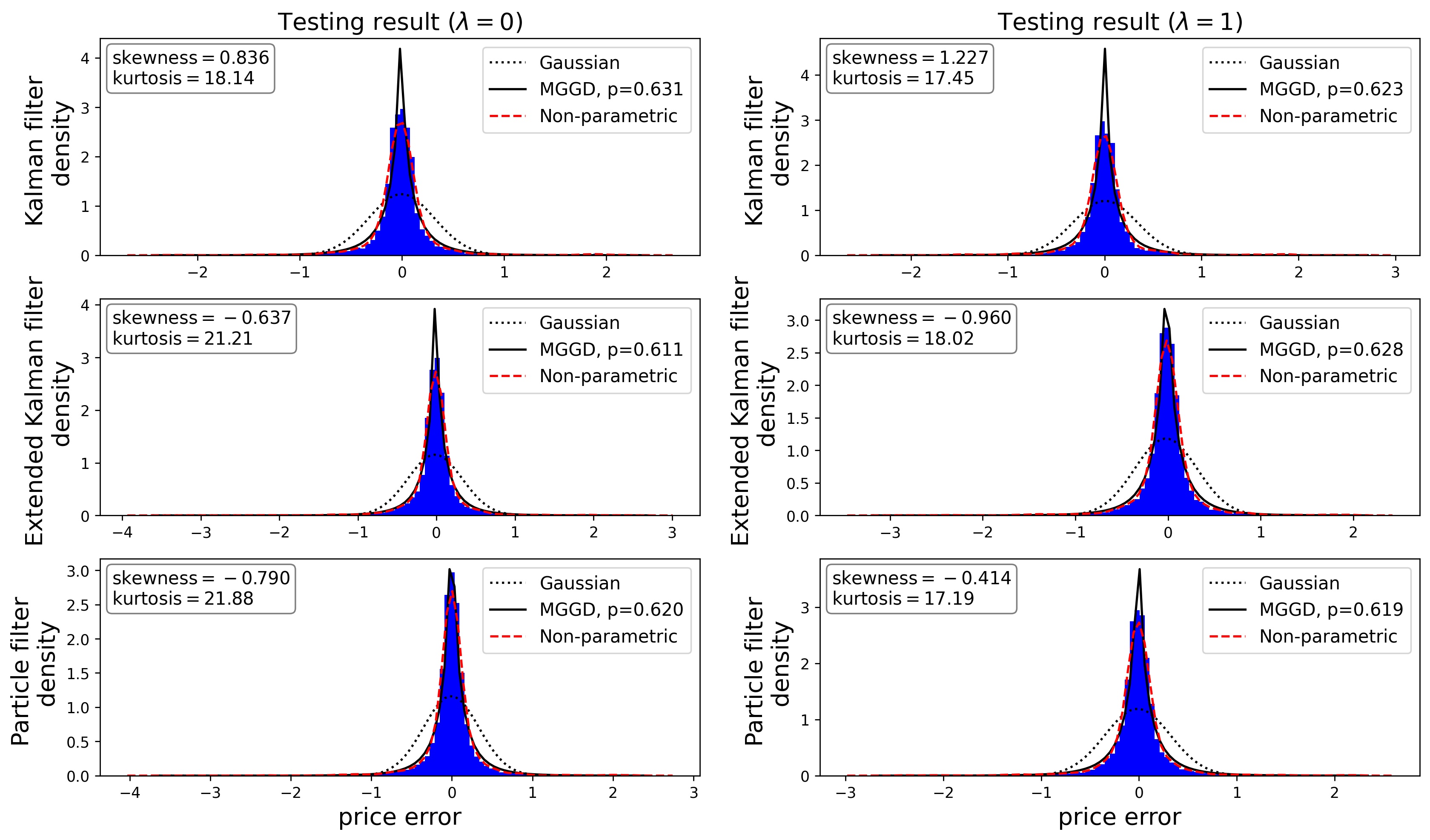}
 		\label{fig6_6_1} %
 	\end{figure}

\paragraph{Particle filter settings.}
Forecast accuracy is stable across reasonable particle counts; runtime scales with particle number and resampling schedule. The effective sample size (ESS) shown in Figures \ref{fig6_5_2} and \ref{fig6_5_3} presents the variance of the particles over the maximum number of particles (300). The value of ESS is between 0 and $100\%$ and the threshold in adaptive resampling is usually at $50\%$. In other words, if the ESS is less than half of the total number of particles (ESS < $N/2$), then the particle filter is considered inefficient and resampling is necessary. In our application, we run systematic resampling at every time step instead of an adaptive method and examine the efficiency of the particle filter using the ESS. In Figure \ref{fig6_5_2}, we vary the arbitrage regularization parameter on or off ($\lambda=0,1$) and compare the ESS with MGGD shape parameter $p=1.5$. The small initial value of ESS in the first step is due to the inexact initial particles which are sampled from the sample means of the estimated state variables and are not exactly the accurate initials for the forward rate curve. The result in Figure \ref{fig6_5_2} shows the ESS of the particles stays above $85\%$ in the training set and decays to $60\%$ in the testing set over time, which also indicates that the particle filter does not suffer from serious degeneracy. In Figure \ref{fig6_5_3}, we vary $p$, the shape parameter of the MGGD distribution, and compare the ESS with arbitrage regularization on ($\lambda=1$). We observe that the MGGD with $p=0.62$ has much higher ESS in the later time steps and the ESS is not decaying in the testing set. Later, we show that the optimal value of $p$ is around $0.62$ in the error distribution of the predicted bond prices. Therefore, we conclude that the particle filter with multivariate generalized Gaussian distribution is very efficient and stable for bond prices forecasting in both training and testing data.

 	\section{Conclusion}\label{sec:conclusion}

We develop an arbitrage-aware forecasting framework that combines state-space filters (KF/EKF/PF) with a deep architecture for time-varying parameters and an explicit no-arbitrage training penalty (AER). Empirically, AER delivers the largest gains at short maturities and at the 5-day horizon, improving bid--ask hit rates while keeping MAE/RMSE competitive (Section~5). Methodologically, the approach bridges HJM/AFNS with neural time-series encoders (LSTM/CLSTM) and differentiable filtering modules, using AER both as a training signal and an ex post diagnostic of arbitrage consistency (cf.\ \cite{KratsiosHyndman2020Risks}).

Looking ahead, we see three natural extensions: (i) adapting the framework to post-LIBOR overnight-rate curves with expected jumps (e.g., \cite{FontanaGrbacSchmidt2024_Overnight}); (ii) evaluating data-assisted filters---such as KalmanNet-style estimators---within our framework without altering the rest of the methodology \cite{Revach2022KalmanNet,Dahan2023BayesianKalmanNet,ShlezingerEldar2023}, and attention-based state-estimation methods with theoretical guarantees (e.g., \citet{HorvathKratsiosLimmerYang2025}); and (iii) pairing AER with arbitrage-consistent neural-SDE or other generative models for scenario analysis (e.g., \cite{Cohen02012023}). Because parameters are learned by gradient-based optimization under a no-arbitrage penalty, the method is suitable for day-to-day pricing, model monitoring, and risk-aware short- to medium-horizon forecasts, with the strongest gains observed at the 5-day horizon. In this context, the empirical analysis is intentionally confined to the pre-COVID period 2017--2019 in order to avoid structural breaks associated with the pandemic and subsequent monetary policy regime shifts. While this limits regime coverage, it allows for a controlled evaluation of arbitrage-free regularization under comparatively stable market conditions. Importantly, the proposed methodology is independent of any particular market environment and can be directly applied to more volatile or stressed periods when comparable data become available.  Incorporating term-structure derivatives and macroeconomic features into estimation is also a natural extension for deeper empirical study.

We implemented Kalman, extended Kalman, and particle filters in TensorFlow and integrated a forward--rate term-structure model into a neural encoder to forecast bond prices. The arbitrage-regularization penalty (AER) is embedded in a Nelson--Siegel--type forward-rate specification and calibrated on time series of coupon bonds under a strong no-arbitrage requirement. The forward-rate model can be extended to the four-parameter Svensson family (and higher orders) while retaining the same no-arbitrage regularization.
Our empirical analysis evaluates both yields and prices, with the most pronounced improvements under arbitrage regularization observed in price-based metrics, particularly 5-day-ahead hit rates and dollar-denominated MAE.
The combination of a theoretically motivated term-structure model with a multi-layer neural architecture (LSTM/CLSTM) produces accurate forecasts. The AER penalty quantifies departures from no-arbitrage and serves as an ex post diagnostic in addition to improving fit. We find that incorporating AER \emph{does not inherently degrade} forecast accuracy; rather, its effect depends on horizon and maturity.

For practical use, different filters trade accuracy for computation. A \emph{KF} in yield-space is fastest but requires extracting yields from prices. An \emph{EKF} in price-space offers a good accuracy--runtime balance. A \emph{PF} in price-space with MGGD observation noise and systematic resampling is the most computationally demanding but accommodates non-Gaussian residuals. In our data, the empirical price-error distribution is markedly non-Gaussian; PF performance improves as the MGGD shape approaches Laplace, consistent with heavier-tailed noise.

The arbitrage penalty derived from the forward-rate model is compatible with a range of fixed-income pricing models. Its implementation adds modest computational overhead, and its benefits are most visible in the settings emphasized here. As noted by \citet{christensen2011affine} and \citet{diebold2006forecasting}, no-arbitrage models remain approximations; even when absence of arbitrage is enforced in theory, small violations can appear in empirical implementations. The periodic behavior we observe in AER across maturities suggests uses in portfolio monitoring and risk management; future work can quantify AER further to flag potential over- or under-valuation. Heavy-tail features may also be modeled via mean-reverting dynamics with jumps. The flexibility of our neural encoder and the dynamic parameterization create a platform for studying classical no-arbitrage theory with richer data and extensions (e.g., Svensson curves, derivatives, macroeconomic factors).

These results demonstrate that enforcing financial structure within deep learning architectures can significantly improve the stability, interpretability, and realism of fixed-income forecasts.

        \nocite{XGao-PHDthesis2021}
	\bibliographystyle{abbrvnat}
	\bibliography{mybib,newrefs}

        \appendix
        \section{Particle filtering implementation} \label{app:PF}
The conditional expected value of $X_t$ from the previous state  $X_{t-1}$ given observations $Y_{1:t-1} = y_{1:t-1}$ is denoted as the posterior distribution $p(X_t\left|Y_{1:{t-1}}\right.)$. The calculation of the expectation is estimated by Monte Carlo sampling
 	\begin{equation*}
 		\mathbb{E}^{\mathbb{p}}\left[f(X)\right] = \frac{f(X_1) + f(X_2) +\cdots + f(X_N)}{N}.
 	\end{equation*}
 	In practice, it is difficult to sample from the posterior distribution $p(X_k\left|Y_{1:k}\right.)$. We assume that we can sample from some prior distribution $q(X_k\left|Y_{1:k}\right.)$ called the {\it importance distribution}, then we can estimate the conditional expectation through the following steps
 	\begin{align}\label{conditional_expectation}
 		\mathbb{E}^{\mathbb{p}}\left[f(X_{t}) \left| Y_{1:t} \right.\right] =& \int f(X_t)p(X_t|Y_{1:t}) dX_t\nonumber\\
 		=& \int f(X_t)\frac{p(X_t|Y_{1:t})}{q(X_t|Y_{1:t})} q(X_t|Y_{1:t}) dX_t\nonumber\\
 		=& \frac{1}{p(Y_{1:t})} \int f(X_t) \frac{p(Y_{1:t} | X_t)p(X_t)}{q(X_t|Y_{1:t})} q(X_t|Y_{1:t}) dX_t\nonumber\\
 		=& \frac{\int f(X_t) w_t(X_t) q(X_t|Y_{1:t}) dX_t}{\int w_t(X_t) q(X_t|Y_{1:t}) dX_t}\nonumber\\
 		=&\frac{\mathbb{E}^q\left[w_t(X_t) f(X_t) | Y_{1:t}\right]}{\mathbb{E}^q\left[w_t(X_t) | Y_{1:t}\right]},
 	\end{align}
 	where
 	\begin{equation*}
 		w_t(X_t) = \frac{p(Y_{1:t} | X_t)p(X_t)}{q(X_t|Y_{1:t})}.
 	\end{equation*}
 	The calculation of (\ref{conditional_expectation}) can be estimated by sampling $\{X_t^{(i)}\} \sim  q(X_t|Y_{1:t})$ for $i=1,\cdots,N$. That is,
 	\begin{align*}
 		\mathbb{E}^{\mathbb{Q}}\left[f(X_{t}) \left| Y_{1:t} \right.\right] =& \frac{\mathbb{E}^q\left[w_t(X_t) f(X_t) \left| Y_{1:t}\right.\right]}{\mathbb{E}^q\left[w_t(X_t) \left| Y_{1:t}\right.\right]}\\
 		=&\frac{\frac{1}{N} \sum_{i=1}^N w_t(X_t^{(i)}) f(X_t^{(i)})}{\frac{1}{N} \sum_{i=1}^N w_t(X_t^{(i)})}\\
 		=& \sum_{i=1}^N \hat w_t(X_t^{(i)}) f(X_t^{(i)}),
 	\end{align*}
 	where $\hat w_t$ are normalized weights
 	\begin{equation*}
 		\hat w_t(X_t^{(i)}) = \frac{w_t(X_t^{(i)})}{\sum_{i=1}^N w_t(X_t^{(i)})}.
 	\end{equation*}
 	
 	Suppose the prior distribution $q(\cdot)$ satisfies the Markov property, then we can rewrite $w_k$ as a recursive identity
 	\begin{equation*}
 		w_t^{(i)} = w_{t-1}^{(i)} \frac{p\left(Y_t \big| X_t^{(i)}\right) p\left(X_t^{(i)} \big| X_{t-1}^{(i)}\right)}{q\left(X_t^{(i)} \big|X_{t-1}^{(i)}, Y_{1:t}\right)}.
 	\end{equation*}
 	If we choose the prior distribution $q\left(X_t \big|X_{t-1}, Y_{1:t}\right) = p\left(X_t\big| X_{t-1} \right)$ which is also widely used, we obtain the simple recursion
 	\begin{equation*}
 		w_t^{(i)} = w_{t-1}^{(i)} p\left(Y_t \big| X_t^{(i)} \right).
 	\end{equation*}
 	This choice of prior distribution does not incorporate the most recent observations $Y_t$, so it is inefficient. \citet{javaheri2003filtering} propose using the extended Kalman filter to obtain the posterior information from the observations. The following distribution with prior mean $\hat X_{k-1|k-1}$ and posterior covariance $P_{k-1}$ from the extended Kalman filter 
 	\begin{equation*}
 		q\left(X_k \big|X_{k-1}, Y_{1:k}\right) = \mathcal{N} \left(X_k \left| g(\hat X_{k-1|k-1}), P_{k-1}\right.\right),
 	\end{equation*}
 	gives one way to implement the importance sampling in particle filter.
 	
 	Standard importance sampling suffers the variance explosion problem since some particles may have increasingly large weights and others have very small weights. The variance of weights increases exponentially with respect the number of particles. This degeneration problem decreases the effectiveness of particles and increases variance of the weights. To address this problem, a resampling step is introduced into the recursive procedure. The resampling is equivalent to resample each particle in such a way that their offspring $o_t = \left(o^{[1]}_t,\cdots,o^{[N]}_t\right)$ follows a multinomial distribution with parameter vector $\left(N, \hat w_t\right)$ and each particle is distributed with equally probability of $1/N$. The resampled distribution is an unbiased estimation of the original particle distribution. As a consequence, resampling carries the computational efforts to retain the particles in dense probability mass by precluding the particles of low weights with high probability. The most widely-used resampling method is systematic resampling introduced by \citet{kitagawa1996monte} which we introduce in the algorithm.
 	
 	On the other hand, resampling also has disadvantages. There could be the situation that a particle having a low weight could have a high weight at the next time and if this happens then resampling could be wasteful. Another immediate effect of resampling is some extra noise being introduced. One way we need resampling to control variance of weights and one way we do not want introduce additional variance. However, a controlled variance of weights benefits more from the additional variance noise after resampling. In practice, it is more sensible to resample only when the variance of the normalized weights reaches some threshold. The commonly used threshold (see \citet{liu2008monte}) is the Effective Sample Size (ESS)
 	\begin{equation*}
 		\text{ESS} = \left(\sum_{i=1}^{N} \left(\hat w^{(i)}_t\right)^2\right)^{-1}.
 	\end{equation*}
 	The ESS takes values between $1$ and $N$ and resampling is usually done when ESS is below $N/2$. This method is called adaptive resampling. In our application we do not apply the adaptive method but we examine the efficiency of our model  by investigating the ESS after the training.
 	
 	\noindent\rule[-0.25\baselineskip]{1\linewidth}{1pt}
 	\smallskip
 	
 	\centerline{Sequential importance resampling (SIR) particle filter}
 	
 	\begin{itemize}[align=left, leftmargin=*]
 		\item[At time $k=0$]
 		\item[1.] Sample initial $X^{(i)}_0$ from the initial states
 		\begin{equation*}
 			X^{(i)}_{0} = \hat X_0 + \hat P_0 W^{(i)},
 		\end{equation*}
 		where $ P_0 = \hat P^{}_0 \hat P_0^T$ is the prior covariance matrix and $W^{(i)}$ is standard Gaussian random number.
 		\item[2.] Update weights by initial observations and resampling to obtain equally distributed particles $\{X^{(i)}_0, w_0^{(i)} = 1/N\}$.
 		\item[From time $k\geq 1$]
 		\item[1.] Importance sampling:

 		From the measurement and updating equations given by (\ref{EKF1}) and (\ref{EKF2}) in EKF, we obtain the posterior particles along with the posterior covariance
 		\begin{align*}
 			\hat X^{(i)}_{k-1|k-1} =& X^{(i)}_{k-1} + K_{k-1} v^{(i)}_{k-1},\\
 			P^{(i)}_{k-1|k-1} =& P^{(i)}_{k-1} - K_{k-1}  M_{k-1} P^{(i)}_{k-1},
 		\end{align*}
 		then we sample particles from the posterior space
 		\begin{align*}
 			X^{(i)}_k = A_{k-1}^{} \hat X^{(i)}_{k-1|k-1} + D_{k-1}^{} + \sqrt{P^{(i)}_k} W^{(i)},
 		\end{align*}
 		where
 		\begin{align*}
 			P^{(i)}_{k} &= A^T_{k-1} P^{(i)}_{k-1|k-1} A^{}_{k-1} + Q^{}_{k-1},\\
 			v^{i}_{k-1} &= Y^{}_{k-1} -  \hat Y(t_{k-1},~X^{(i)}_{k-1}),\\
 			F^{}_{k-1} &= M^{}_{k-1} P^{(i)}_{k|k-1}  M^T_{k-1} + U_{k-1}^{},\\
 			K^{}_{k-1} &= P^{(i)}_{k-1}  M_{k-1}^T F_{k-1}^{-1},\\
 			M^{}_{k-1} &= \left.\frac{\partial \hat Y}{\partial X}\right|_{X={X}^{(i)}_{k-1}}.
 		\end{align*}
 		
 		\item[2.] Update weights:
 		\begin{equation*}
 			w_k^{(i)} = w^{(i)}_{k-1} \frac{p\left(Y_k \big| X_k^{(i)}\right) p\left(X_k^{(i)} \big| X_{k-1}^{(i)}\right)}{q\left(X_k^{(i)} \big| x_{k-1}^{(i)}, Y^{}_{1:k}\right)},
 		\end{equation*}
 		where
 		\begin{align*}
 			&p\left(Y^{}_k \big|X_k^{(i)}\right) = \mathcal{M}\left(Y^{}_k \left| \hat Y(X_k^{(i)}), U_k^{}\right.\right),\\
 			&p\left(X_k^{(i)} \big| X_{k-1}^{(i)}\right) = \mathcal{N}\left(X_k^{(i)} \left|  g(X^{(i)}_{k-1}), Q^{}_{k-1} \right.\right),\\
 			&q\left(X_k^{(i)} \big| X_{k-1}^{(i)}, Y_{1:k}\right) = \mathcal{N}\left(X_k^{(i)} \left|  g(X^{(i)}_{k-1|k-1}), P^{(i)}_{k}\right.\right).
 		\end{align*}
 		Calculate normalized weights
 		\begin{equation*}
 			\bar w_k^{(i)} = \frac{w_k^{(i)}}{\sum_{i=1}^N w_k^{(i)}}.
 		\end{equation*}
 		\item[3.] Systematic Resampling from $\left\{\bar{w}^{(i)}_k, X^{(i)}_k, P^{(i)}_k\right\}$ to obtain equally weighted particle sample $\left\{\frac{1}{N}, X^{(i)}_k, P^{(i)}_k\right\}$
 		\begin{itemize}[leftmargin=*]
 			\item[i.] Set $s^{(i)}_k = \frac{i-1 + \tilde s_k}{N}$ with $\tilde s_k\sim \mathcal{U}[0,1]$ for $i=1,\cdots,N$.
 			\item[ii.] Then set the number of particles equal to the offspring
 			\begin{equation*}
 				o^{(i)}_k = \left|\left\{s^{(j)}_k:\sum_{n=1}^{i-1} \bar w^{(n)}_k \leq s^{(j)}_k \leq \sum_{n=1}^{i} \bar w^{(n)}_k\right\}\right|,
 			\end{equation*}
 			which is the number of $s^{(j)}_k$ that locates in $\left[\sum_{n=1}^{i-1} \bar w^{(n)}_k , \sum_{n=1}^{i} \bar w^{(n)}_k\right]$.
 		\end{itemize}

 	\end{itemize}		
 	\noindent\rule[-0.25\baselineskip]{1\linewidth}{1pt}

        \section{LSTM architecture} \label{app:LSTM}

In this appendix we follow the  standard formulation of \citet{hochreiter1997long}. The LSTM block takes as inputs $x$ and the output $c_{t-1}$ and hidden state $h_{t-1}$ from previous LSTM block and generates updated values of $c_{t}$ and $h_{t}$ correspondingly:
 	\begin{equation*}
 		(c_{t},  h_{t}) = L(x, (c_{t-1}, h_{t-1})).
 	\end{equation*}
 The  equations of the LSTM are composed of the following four dense layers each serving a different purpose
 	\begin{align}\label{eq:LSTM}
 		f_t &=  a_g\left(\mathpzc{W}_f x + \mathpzc{W}'_f h_{t-1} + \mathpzc{b}_f\right),\nonumber\\
 		i_t &=  a_g\left(\mathpzc{W}_i x + \mathpzc{W}'_i h_{t-1} + \mathpzc{b}_i\right),\nonumber\\
 		o_t &=  a_g\left(\mathpzc{W}_o x + \mathpzc{W}'_o h_{t-1} + \mathpzc{b}_o\right)\\
 		\tilde c_t & =  a_c\left(\mathpzc{W}_c x + \mathpzc{W}'_c h_{t-1} + \mathpzc{b}_c\right),\nonumber\\
 		c_t & =  f_t \circ c_{t-1} + i_t\circ \tilde c_t,\nonumber\\
 		h_t & =  o_t \circ a_h\left(c_t\right),\nonumber
 	\end{align}
 	where the operator $\circ$ denotes the Hadamard product (element-wise product). In each LSTM cell, we have four gates (or layers):
        $f_{t}$, $i_{t}$, $o_{t}$ are the forget, input, and output gates; $h_{t}$ is the hidden layer; $c_{t}$ is the cell state.  The activation functions used are
 	\begin{align*}
 		a_g(x) &= \frac{1}{1 + e^{-x}},\\
 		a_c(x) &= \tanh(x),\\
 		a_h(x) &= \tanh(x).
 	\end{align*}

For  $X_t\in\mathbb{R}^{N}$ with feature size $N$ and the predefined hidden units $H$, the model weights and biases are predefined by
 	\begin{equation*}
 		\mathpzc{W}, \mathpzc{W}' \in \mathbb{R}^{H\times N},\text{ and }\mathpzc{b} \in \mathbb{R}^{H}.		
 	\end{equation*}
The input layer of the linear model (yield-space) with two connected LSTM $L_1$ and $L_2$ at time step $t$ with input $Y_t$ can be written as
 	\begin{align*}
 		\left(c^{I_1}_t, h^{I_1}_t\right) &= L_1\left(Y_t, \left(c^{I_1}_{t-1}, h^{I_1}_{t-1}\right)\right),\\
 		\left(c^{I_2}_t, h^{I_2}_t\right) &= L_2\left(c^{I_1}_t, \left(c^{I_2}_{t-1}, h^{I_2}_{t-1}\right)\right),
 	\end{align*}
where $c^I_t = c^{I_2}_t$ is the output from input layer.

        \section{CLSTM architecture}\label{app:CLSTM}

The nonlinear model (price-space) is trained with the price data in a $4$-dimensional tensor with a size of $S\times T\times N\times F$ where $N$ is the number of bonds and $F=4$ is the feature size. The input data at each time step is then a $N\times F$ matrix that cannot be fed into a standard LSTM.
Hence, we apply a convolutional-LSTM (CLSTM) to decrease the input dimension from $N\times F$ to $1\times H$ for an integer hyperparameter $H$. We then connect it to the standard LSTM. The CLSTM is usually applied for image processing but we can consider our input as an image of single channel $N\times F\times 1$ and use the convolution operation to obtain a vector output of any size $H$. The compact forms of equations of the CLSTM are similar to the standard LSTM (\ref{eq:LSTM}) but using convolution instead of matrix product

We first apply a \textbf{convolutional--LSTM (CLSTM)} to compress per-time-step cross-sectional features into $H$ channels, then feed the result into an LSTM for temporal dynamics.
 	\begin{align*}
 		f_t = & a_g\left(\mathpzc{W}_f \ast x + \mathpzc{W}'_f \ast h_{t-1} + \mathpzc{b}_f\right),\nonumber\\
 		i_t = & a_g\left(\mathpzc{W}_i \ast x + \mathpzc{W}'_i \ast h_{t-1} + \mathpzc{b}_i\right),\nonumber\\
 		o_t = & a_g\left(\mathpzc{W}_o \ast x + \mathpzc{W}'_o \ast h_{t-1} + \mathpzc{b}_o\right),\\
 		\tilde c_t = & a_c\left(\mathpzc{W}_c \ast x + \mathpzc{W}'_c \ast h_{t-1} + \mathpzc{b}_c\right),\nonumber\\
 		c_t = & f_t \circ c_{t-1} + i_t\circ \tilde c_t ,\nonumber\\
 		h_t = & o_t \circ a_h\left(c_t\right),\nonumber
 	\end{align*}
 	where the operator $ (\ast) $ denotes the convolution operation. The kernel of the convolution LSTM is defined by
 	\begin{equation*}
 		\mathpzc{W},\mathpzc{W}' \in \mathbb{R}^{K_W\times K_H\times K_D},
 	\end{equation*}
 	with $K_W$ as width, $K_H$ as height and $K_D$ as depth. The convolution of $\mathcal{W}\in\mathbb{R}^{K_W\times K_H\times K_D}$ and $x\in\mathbb{R}^{N\times F\times1}$ is a tensor of dimension
 	\begin{equation*}
 		\text{dim}\left(\mathpzc{W} \ast x\right) = \left(\left\lfloor\frac{N + 2p - K_W}{s_W} + 1\right\rfloor, \left\lfloor\frac{F + 2p - K_H}{s_H} + 1\right\rfloor, K_D\right),
 	\end{equation*}
 	where $p$ is the size of padding typically set to $0$, $\left(s_W,s_H\right)$ is the size of stride, and the operator $\lfloor\cdot\rfloor$ takes the integer part. To reduce the dimension of the input and obtain $H$-dimensional vector output, we set stride size $(s_W,s_H)=(1,1)$ and kernel size $\left(K_W,K_H, K_D\right)=\left(\lfloor\frac{N}{H}\rfloor,F,1\right)$ for some hyper-parameters $H< N$, and eventually obtain an output with a size of $(H,1,1)$ which can be compressed to $1\times H$. We then connect it to a standard LSTM. The input layer for the nonlinear model (price-space) consisting of a CLSTM $L_c$ and a standard LSTM $L$ is given by
 	\begin{align*}
 		Y_t \in\mathbb{R}^{N\times F}&:\rightarrow Y_t' \in\mathbb{R}^{N\times F\times 1},\\
 		\left(c^{I_c}_t, h^{I_c}_t\right) =& L_c\left(Y'_t, \left(c^{I_c}_{t-1}, h^{I_c}_{t-1}\right)\right),\\
 		c^{I_c}_t \in\mathbb{R}^{H\times 1\times1}&:\rightarrow c'_t \in\mathbb{R}^{1\times H},\\
 		\left(c^{I}_t, h^{I}_t\right) =& L\left(c'_t, \left(c^{I}_{t-1}, h^{I}_{t-1}\right)\right).
 	\end{align*}

\end{document}